\catcode`\@=11

\newif\if@fewtab\@fewtabtrue

{\count255=\time\divide\count255 by 60
\xdef\hourmin{\number\count255}
\multiply\count255 by-60\advance\count255 by\time
\xdef\hourmin{\hourmin:\ifnum\count255<10 0\fi\the\count255}}
\def\ps@draft{\let\@mkboth\@gobbletwo
    \def\@oddhead{}
    \def\@oddfoot
       {\hbox to 7 cm{$\scriptstyle Draft\ version:\ \draftdate$
       \hfil}\hskip -7cm\hfil\rm\thepage \hfil}
    \def\@evenhead{}\let\@evenfoot\@oddfoot}

\def\ceqno{\global\@fewtabfalse
    \ifcase\@eqcnt \def\@tempa{& & &}\or \def\@tempa{& &}
      \or \def\@tempa{&}
      \or\def\@tempa{}\fi\@tempa
{\rm(\theequation)}}

\def\aeqno#1{\global\@fewtabfalse
    \ifcase\@eqcnt \def\@tempa{& & &}\or \def\@tempa{& &}
      \or \def\@tempa{&}
      \or\def\@tempa{}\fi\@tempa
{\rm(\theequation,#1)}}

\def\label#1{\ifnum\draftcontrol=1
 \global\def\draftnote{$\scriptstyle #1$}\fi
 \@bsphack\if@filesw {\let\thepage\relax
   \def\protect{\noexpand\noexpand\noexpand}%
\xdef\@gtempa{\write\@auxout{\string
      \newlabel{#1}{{\@currentlabel}{\thepage}}}}}\@gtempa
   \if@nobreak \ifvmode\nobreak\fi\fi\fi
  \@esphack}

\def\alabel#1#2{\label{#1}\global\@fewtabfalse
    \ifcase\@eqcnt \def\@tempa{& & &}\or \def\@tempa{& &}
      \or \def\@tempa{&}
      \or\def\@tempa{}\fi\@tempa
{\hbox to 3cm{\phantom{\rm(\theequation,#2)}
\draftnote \hfil}\hskip -3cm {\rm(\theequation,#2)}}}

\def\clabel#1{\label{#1}\global\@fewtabfalse
    \ifcase\@eqcnt \def\@tempa{& & &}\or \def\@tempa{& &}
      \or \def\@tempa{&}
      \or\def\@tempa{}\fi\@tempa
{\hbox to 3cm{\phantom{\rm(\theequation)}
\draftnote \hfil}\hskip -3cm{\rm(\theequation)}}}

\def\eqnarray{\def\draftnote{{}}\global\@fewtabtrue
\stepcounter{equation}\let\@currentlabel=\theequation
\global\@eqnswtrue
\global\@eqcnt\z@\tabskip\@centering\let\\=\@eqncr
$$\halign to \displaywidth\bgroup\@eqnsel\hskip\@centering\@eqcnt\z@
  $\displaystyle\tabskip\z@{##}$&\global\@eqcnt\@ne
  \hskip 1\arraycolsep \hfil${##}$\hfil
  &\global\@eqcnt\tw@ \hskip 1\arraycolsep
$\displaystyle\tabskip\z@{##}$
\hfil  \tabskip\@centering&\global\@eqcnt\thr@@\llap{##}\tabskip\z@
\cr}

\def\endeqnarray{\@@eqncr\egroup
      \global\advance\c@equation\m@ne$$\global\@ignoretrue}

\def\@eqnnum{\hbox to 3cm{\phantom{\rm(\theequation)} \draftnote
                         \hfil}\hskip -3cm {\rm(\theequation)}}

\def\@@eqncr{\let\@tempa\relax
    \ifcase\@eqcnt \def\@tempa{& & &}\or \def\@tempa{& &}
      \or \def\@tempa{&}
      \or\def\@tempa{}
\fi\@tempa
\if@eqnsw
\if@fewtab\@eqnnum\fi
\stepcounter{equation}\fi\global
\@eqnswtrue\global\@eqcnt\z@\global\@fewtabtrue\cr}

\def\draftcite#1{\ifnum\draftcontrol=1#1\else{}\fi}

\def\@lbibitem[#1]#2{\item{}\hskip -3cm \hbox to 2cm
{\hfil$\scriptstyle\draftcite{#2}$}\hskip
1cm[\@biblabel{#1}]\if@filesw
     {\def\protect##1{\string ##1\space}\immediate
      \write\@auxout{\string\bibcite{#2}{#1}}}\fi\ignorespaces}

\def\@bibitem#1{\item\hskip -3cm \hbox to 2cm
{\hfil $\scriptstyle\draftcite{#1}$}\hskip 1cm
\if@filesw \immediate\write\@auxout
       {\string\bibcite{#1}{\the\value{\@listctr}}}\fi\ignorespaces}

\def\nsection#1{\section{#1}\setcounter{equation}{0}}
\font\tendl=msbm10  scaled \magstep1%double line
\font\sevendl=msbm7 scaled \magstep1
\font\fivedl=msbm5 scaled \magstep1
\font\tengl=eufm10  scaled \magstep1% gothic letters
\font\sevengl=eufm7 scaled \magstep1
\font\fivegl=eufm5 scaled \magstep1

\newfam\dlfam  % \dl is double line
\textfont\dlfam=\tendl \scriptfont\dlfam=\sevendl
\scriptscriptfont\dlfam=\fivedl
\newfam\glfam  % \gl is gothic letters
\textfont\glfam=\tengl \scriptfont\glfam=\sevengl
\scriptscriptfont\glfam=\fivegl

\def\draftdate{\number\month/\number\day/\number\year\ \ \ \hourmin }

\global\def\draftcontrol{0}
\catcode`\@=12
\def\tilde{\widetilde}
\def\hat{\widehat}
\documentclass[aps,superscriptaddress,floatfix]{revtex4}
\usepackage{epsfig}
\usepackage[centerlast]{subfigure}
\usepackage{bm}
\usepackage{amssymb}

\renewcommand{\theequation}{\thesection.\arabic{equation}}

\newcommand{\be}{\begin{eqnarray}}
\newcommand{\en}{\end{eqnarray}\vs 0.5 cm}

\newcommand{\Id}{{I\hspace{-0.04cm}d}}
\newcommand{\Gr}{{G\hspace{-0.04cm}r}}

\newcommand{\vs}{\vskip}

\newcommand{\NR}{{{\bf R}}}%letra doble raya en modo matematico
\newcommand{\qq}{\begin{eqnarray}}

\newcommand{\ee}{{\rm e}}

\newcommand{\qqq}{\end{eqnarray}}

\newcommand{\tr}{\hbox{tr}}

\newcommand{\CC}{{\cal C}}
\newcommand{\CD}{{\cal D}}
\newcommand{\CE}{{\cal E}}

\newcommand{\CH}{{\cal H}}

\newcommand{\CJ}{{\cal J}}

\newcommand{\CL}{{\cal L}}

\newcommand{\CO}{{\cal O}}
\newcommand{\CP}{{\cal P}}

\newcommand{\m}{\hspace{0.025cm}}

\begin{document}
\title{\Large\bf{Kraichnan flow in a square:\\ 
an example of integrable chaos}}
\author{Rapha$\rm\ddot{e}$l Chetrite} \affiliation{Laboratoire de Physique, 
ENS-Lyon, 46 All\'ee d'Italie, 69364 Lyon, France}
\author{Jean-Yves Delannoy} \affiliation{Laboratoire de Physique, 
ENS-Lyon, 46 All\'ee d'Italie, 69364 Lyon, France}
\author{Krzysztof Gaw\c{e}dzki}\affiliation{Laboratoire de Physique, 
ENS-Lyon, 46 All\'ee d'Italie, 69364 Lyon, France} \affiliation{member of C.N.R.S.}

\begin{abstract}

\noindent The Kraichnan flow provides an example of a random dynamical system
accessible to an exact analysis. We study the evolution of the infinitesimal 
separation between two Lagrangian trajectories of the flow. Its long-time 
asymptotics is reflected in the large deviation regime of the statistics 
of stretching exponents. Whereas in the flow that is isotropic at small 
scales the distribution of such multiplicative large deviations is Gaussian, 
this does not have to be the case in the presence of an anisotropy. We 
analyze in detail the flow in a two-dimensional periodic square where the 
anisotropy generally persists at small scales. The calculation of the large 
deviation rate function of the stretching exponents reduces in this case 
to the study of the ground state energy of an integrable periodic 
Schr\"{o}dinger operator of the Lam\'e type. The underlying integrability 
permits to explicitly exhibit the non-Gaussianity of the multiplicative 
large deviations and to analyze the time-scales at which the large deviation 
regime sets in. In particular, we indicate how the divergence of some of 
those time scales when the two Lyapunov exponents become close allows 
a discontinuity of the large deviation rate function in the parameters 
of the flow. The analysis of the two-dimensional anisotropic flow permits 
to identify the general scenario for the appearance of multiplicative 
large deviations together with the restrictions on its applicability.
\end{abstract}
\maketitle
\nsection{Introduction}

\noindent The Kraichnan random ensemble of velocities \cite{Kr68} has been 
extensively used to model various phenomena related to turbulent transport 
both in the inertial interval of scales that develops at high Reynolds 
numbers and at moderate Reynolds numbers where the viscosity effects play 
an important role \cite{FGV}. The passive transport of scalar or vector 
quantities in a velocity field is governed by the Lagrangian flow describing
the evolution of the trajectories of fluid particles. From the mathematical
point of view, such flow provides an example of a random dynamical system  
\cite{LArnold} which in the case of the Kraichnan velocities is described by a 
stochastic differential equation. For the Kraichnan flows corresponding to 
moderate Reynolds numbers, the methods borrowed from the theory of random 
dynamical systems or stochastic differential equations appear to provide 
important information about the transport properties of the flows. To start 
with, the values of the Lyapunov exponents of the flow, whose existence is 
asserted by the multiplicative ergodic theorem, allow to decide whether
the flow is chaotic (positive top Lyapunov exponent) or not, leading to 
different directions of the cascades of passively advected scalars \cite{CKV}.
More detailed information about the transport properties of the flow
may be extracted from the knowledge of the fluctuations of the exponential 
stretching rates around their limiting long-time values equal to the Lyapunov
exponents. In the generic case where all the Lyapunov exponents are different,
the statistics of the stretching exponents may be expected to exhibit at long 
but finite times a large deviation regime captured quantitatively by a single
function of the vector of the stretching rates. Since the existence of such 
multiplicative large deviation regime is not assured by general mathematical 
theorems, see \cite{BaxStr} for partial results, it is interesting to have
at our disposal models where it may be established and studied in detail. 
\vskip 0.1cm

One such example that has been known for some time is the homogeneous isotropic
Kraichnan flow. The corresponding stochastic differential equation has been
studied in the mathematical literature in the eighties and nineties of the
last century. In particular, the Lyapunov exponents have been found in
\cite{LJ} and \cite{Bax} 
%and appeared to be equidistant 
but the large deviations have been studied only separately for the top stretching 
exponent or for the sum of them \cite{BaxStr}. With the regain of 
interest of physicists in the Kraichnan model in the mid-nineties, 
the same stochastic equation resurfaced as the model for the Lagrangian flow 
at moderate Reynolds numbers with the motivations, the accents and the language 
proper to the turbulence theory community \cite{CFKL}. In particular, it was 
realized that many properties of the turbulent transport require more information
about the flow than the spectrum of Lyapunov exponents and may be expressed 
in terms of the rate function of the large deviations of the stretching 
exponents. Those include the rate of decay of the moments of advected scalar 
\cite{BF} or of growth of those of the magnetic field \cite{CFKV} or, in 
compressible flows, of the density fluctuations \cite{BFF}, the multifractal 
properties of long-time density concentrations \cite{BGH} and the threshold
for the onset of the drag reduction in polymer solutions \cite{Chert}.
In the homogeneous and isotropic Kraichan flow, the large deviation regime
of the stretching exponents is Gaussian and the corresponding rate
function is a quadratic polynomial \cite{BF,BFF}. Its simple form has permitted 
to extract analytic answers for many characteristics features of passive 
advection in such flows \cite{FGV}. 
\vskip 0.1cm

The simplicity of the multiplicative large deviation regime in the homogeneous 
isotropic Kraichnan flow is due to the decoupling of the dynamics of the
stretching exponents from that of the eigen-directions for stretching and
contraction \cite{BF}. As a result also the exact distribution of the 
stretching exponents may be found analytically in this case as it appears to 
be related to the heat kernel of the integrable quantum Calogero-Sutherland 
Hamiltonian for particles on the line interacting with the attractive pair 
potential proportional to the function $\,\sinh^{-2}\,$ of the inter-particle
distance \cite{SM,CIRM}. In the present paper we analyze the two-dimensional 
Kraichnan flow in a periodic square often used in numerical simulations. The 
large scale anisotropy due to the shape of the flow volume generically persists 
on small scales inducing isotropy breaking terms in the distribution of strain 
that drives the evolution of the stretching exponents. Due to the presence of such 
terms, the stretching exponents dynamics does not decouple anymore from that 
of the (unstable) eigen-directions. The Lyapunov exponents may nevertheless be 
still computed analytically and their difference expressed in terms of elliptic 
integrals. The distribution of the sum of the stretching exponents is still Gaussian 
for all times (this is a general fact for the homogeneous Kraichnan flows). As for 
the rate function of the large deviations of the difference of the stretching exponents, 
its calculation may be reduced to that of the ground-state energy of the integrable 
one-dimensional periodic quantum Lam\'e operator. For general values of the coupling 
constant, the eigenvalues of the latter may be found by numerical diagonalization 
of infinite tridiagonal matrices \cite{ErdMOT}. Those matrices reduce to finite ones 
at integer values of the coupling and for the lowest eigenvalues. Alternatively, 
the ground state energy of the Lam\'e operator may be found by direct numerical 
integration of the eigenvalue equation. Both approaches permit to obtain the large 
deviation rate function for the stretching exponents that turns out to be non-quadratic 
although with quadratic asymptotes. The analysis of the spectral gap
of the Lam\'e operator permits to assess the time scales at which the 
multiplicative large deviation regime sets in. In particular, the divergence 
of the time scales relative to the multiplicative central-limit regime when 
the difference of the Lyapunov exponents tends to zero accompanies the 
observed discontinuity of the large deviation rate function in the 
anisotropy parameter at the point where the Lyapunov exponents coincide. 
\vskip 0.1cm

The plan of the paper is as follows. In Sect.\,II, we discuss the
relations between the Lagrangian flow and random dynamical systems
introducing the concepts of the natural invariant measure and of 
the tangent process and stating two different definitions of the 
stretching exponents. In Sect.\,III we introduce the Kraichnan ensemble 
of velocities and discuss how the general concepts considered before 
simplify in the homogeneous Kraichnan velocities. We recall briefly the 
results about the statistics of the stretching exponents in the isotropic
version of the Kraichnan model. Sect.\,IV is the core of the present 
paper. We discuss there the Kraichnan flow in a periodic square, 
the persistence of anisotropy at small distances, the calculation
of the Lyapunov exponents and, finally, the large deviations
for the stretching exponents, their relation to the Lam\'e equation, 
their non-Gaussianity, and their discontinuity in the anisotropy 
parameters. The last section collects our conclusions. We believe that 
the simple model considered here allows to identify a general scenario 
for the occurrence of the multiplicative large deviations when all 
the Lyapunov exponents are different and to understand the mechanism 
of its failure when some of the Lyapunov exponents get close.       
\vskip 0.3cm

\noindent{\bf Acknowledgements}. \ The work of R.C. on the present project
was started during his stay as trainee at the Department of Physics of 
Complex Systems at Weizmann Institute in Rehovot. The research of K.G. was 
partially done in framework of the European contracts Stirring and 
Mixing/HPRN-CT-2002-00300 and Euclid/HPRN-CT-2002-00325. R.C. and K.G. 
thank Grisha Falkovich and Sasha Fouxon for discussions that initiated 
this work. K.G. acknowledges discussions with Giovanni Gallavotti and 
helpful comments about Lam\'e operator from Giovanni Felder, Edwin Langmann 
and, especially those from Hans Volkmer who made available to us his notes 
and a relevant Maple program.

\nsection{Lagrangian flow as a random dynamical system}
\subsection{Random dynamical systems}

\noindent We shall start by describing the Lagrangian flow in the language of
random dynamical systems. Let us consider an ensemble of velocities 
$\,\bm u^\omega_t(\bm r)\,$ in a bounded region $\,V\,$ of the $d$-dimensional 
space. Here $\,\omega\,$ is a random parameter belonging to a probability 
space $\,\Omega\,$ equipped with a probability measure $\,P(d\omega)$. 
$\,\Omega\,$ may be taken as the space of velocity realizations. We shall 
assume the stationarity of the velocity ensemble, i.e. the existence of 
a 1-parameter group of measure preserving transformations $\,\omega\mapsto
\omega_s$ of $\,\Omega\,$ such that $\,\bm u^{\omega_{s}}_t(\bm r)
=\bm u^\omega_{t+s}(\bm r)$. \,The Lagrangian flow describing the 
trajectories of tracer particles carried by the fluid is defined by the 
ordinary differential equation
\qq  
{d{\bm R}\over dt}\ =\ {\bm u}^\omega_t({\bm R})\,.
\label{Lf}
\qqq
Under simple regularity assumptions including the 
spatial smoothness of the velocities  $\,\bm u^\omega_t(\bm r)$, 
\,the solutions $\,\NR^\omega_t(\bm r)\,$
of Eq.\,(\ref{Lf}) parametrized by their time zero position 
$\,\bm r\,$ define a family of random smooth maps $\,\Phi^\omega_t\,$ 
of the region $\,V\,$ such that $\,\NR^\omega_t(\bm r)
=\Phi^\omega_t(\bm r)\,$ with the composition rule
\qq
{\bm\Phi}^\omega_{s+t}\ =\ {\bm \Phi}^{\omega_{s}}_t\circ{\bm\Phi}^\omega_s\,.
\label{crl}
\qqq
In particular, one obtains a 1-parameter group of transformations
\qq
(\bm r,\,\omega)\ \longmapsto\ (\Phi^\omega_t(\bm r),\,\omega_{t})
\label{1pg}
\qqq
of the product space $\,V\times\Omega\,$ which realizes the flow dynamics. 

\subsection{Natural invariant measure}
\label{sec:nim}

\noindent Note that $\,\Phi^{\omega_{s}}_{-s}(\bm r)\,$ is the time
zero position of the solution that at time $\,s\,$ passes through $\,\bm r$.
\,Suppose that, for continuous functions $\,f\,$ on $\,V$,
\,the limit 
\qq
\lim\limits_{s\to-\infty}\,\,{1\over|V|}\int\limits_V f(\Phi^{\omega_{s}}
_{-s}(\bm r))\,d\bm r\ =:\ \int\limits_V f(\bm r)\,\mu^\omega(d\bm r)
\label{mu}
\qqq
exists for almost all $\,\omega\,$ and defines a random family
of probability measures $\,\mu^\omega(d\bm r)\,$ on $\,V$. 
\,Note that due to the composition rule (\ref{crl}),
\qq
\int\limits_V f(\Phi^\omega_t(\bm r))\,\mu^\omega(d\bm r)\ =\ 
\int\limits_V f(\bm r)\,\mu^{\omega_{t}}(d\bm r)\,.
\label{mdy}
\qqq
The measures $\,\mu^\omega(d\bm r)\,$ describe the distribution of the time 
zero positions of the Lagrangian trajectories whose initial 
points were uniformly seeded in the far past. For incompressible 
velocities the uniform distribution
is conserved by the flow so that $\,\mu^\omega(d\bm r)={1\over|V|}d\bm r\,$ 
but in the presence of compressibility, the Lagrangian trajectories develop
preferential concentrations and the measures $\,\mu^\omega(d\bm r)\,$
tend to be singular and supported by lower dimensional random attractors. 
\vskip 0.3cm

From the random measures $\,\mu^\omega(d\bm r)\,$ one may synthesize 
a measure $\,M(d\bm r,d\omega)\,$ on the product space $\,V\times\Omega\,$ 
defined by the relation
\qq
\int\limits_{V\times\Omega}f(\bm r,\omega)\,M(d\bm r,d\omega)\ =\ 
\bigg\langle\int\limits_Vf(\bm r,\omega)\,\mu^\omega(d\bm r)\bigg\rangle
\nonumber
\qqq
where $\,\Big\langle\cdot\cdot\,\Big\rangle\,$ denotes the expectation w.r.t.
the probability measure $\,P(d\omega)$. 
\,Note that due to the property (\ref{mdy}),
\qq
\int\limits_{V\times\Omega}f(\Phi^\omega_t(\bm r),\omega_t)\,M(d\bm r,d\omega)
\ =\ \int\limits_{V\times\Omega}f(\bm r,\omega)\,M(d\bm r,d\omega)
\nonumber
\qqq
so that the measure $\,M(d\bm r,d\omega)\,$ is invariant under 
the 1-parameter group (\ref{1pg}) of dynamical transformations. 
We shall call $\,M(d\bm r,d\omega)\,$ the {\bf natural invariant measure} 
of the random dynamical system (\ref{Lf}). Below, we shall assume that 
$\,M(d\bm r,d\omega)\,$ is ergodic with respect to the group 
(\ref{1pg}), i.e. that functions invariant under the dynamics 
are constant $\,M$-almost everywhere. 

\subsection{Tangent process}

\noindent Much information about the Lagrangian flow may be extracted 
by looking at the evolution of the separation between two infinitesimally 
close trajectories. Consider the Jacobi matrix 
$\,\nabla\Phi^\omega_t(\bm r)\equiv W^\omega_t(\bm r)\,$ with 
the entries
\qq
W^i_{\,\,j}\ =\ \nabla_{j}{\Phi^\omega_t}^i(\bm r)\,.
\nonumber
\qqq
The matrix $\,W^\omega_t(\bm r)\,$ propagates the infinitesimal separations:
\qq
\delta\bm R^\omega(t;{\bm r})\ =\ W^\omega_t({\bm r})\,\delta\bm r\,.
\nonumber
\qqq
Note that $\,W^\omega_0(\bm r)=\Id\,$ and that $\,W^\omega_t(\bm r)\,$ 
satisfies the linear differential equation
\qq
{dW\over dt}\ =\ S^\omega_t\,W
\label{lde}
\qqq
with $\,(S^\omega_t)^i_{\,\,j}=\nabla_j
{u^\omega_t}^i(\bm R^\omega_t(\bm r))\,$
equal to the matrix elements of the strain along the Lagrangian
trajectory. \,We shall call $\,W^\omega_t(\bm r)\,$ the 
{\bf tangent process}. In studying it below, Eq.\,(\ref{lde}) will play 
a crucial role. Note that
the composition rule (\ref{crl}) implies that
\qq
W^\omega_{s+t}(\bm r)\ =\ W^{\omega_s}_t(\Phi_s^\omega(\bm r))\,
W^\omega_s(\bm r)\,.
\nonumber
\qqq
In particular,
\qq
W^\omega_{-t}(\bm r)\ =\ W^{\omega_{-t}}_t(\Phi_{-t}^\omega(\bm r))^{-1}\,.
\label{invs}
\qqq
\vskip 0.2cm

We shall be interested in the statistical properties of the 
$\,d\times d\,$ matrices $\,W^\omega_t(\bm r)\,$ for fixed but large 
$\,t\,$ with $\,(\bm r,\omega)\,$ sampled according to the natural invariant 
measure $\,M(d\bm r,d\omega)$. \,As for any real invertible $\,d\times d\,$ 
matrix, one may decompose
\qq
W\ =\ O'\,\,{\rm diag}[\ee^{\rho_1},\dots,\ee^{\rho_d}]\,\,O
\label{stre}
\qqq
with a diagonal positive definite matrix sandwiched in between orthogonal
ones. One may demand that the {\bf stretching exponents} $\,\rho_i
={\rho^\omega_i}_{\hspace{-0.06cm}t}(\bm r)\,$ given by half the logarithm
of the eigenvalues of the matrices $\,\ln{W^TW}\,$ and $\,\ln{WW^T}$, 
\,be ordered so that $\,\rho_1\geq\cdots\geq\rho_d$. \,They carry an 
important part of the information about the tangent process. The joint 
probability distribution function (PDF) of the time $\,t\,$ stretching 
exponents is given by the formula:
\qq
P_t(\bm\rho)\ =\ \int\limits_{V\times\Omega}\prod\limits_{i=1}^d\delta(\rho_i
-{\rho_i^\omega}_{\hspace{-0.06cm}t}(\bm r))\,M(d\bm r,d\omega)\,.
\nonumber
\qqq
Note the relation  
\qq
{\rho_i^\omega}_{\hspace{-0.05cm}-t}(\bm r)\ =\ -{\rho_{d-i+1}^{\omega_{{-t}}}}_t
(\Phi_{-t}^\omega(\bm r))
\nonumber
\qqq
between the forward and the backward exponents that follows from 
Eq.\,(\ref{invs}). The invariance of the natural measure under 
the 1-parameter dynamics (\ref{1pg}) implies then that
\qq
P_{-t}(\rho_1,\dots,\rho_d)\ =\ P_{t}(-\rho_d,\dots,-\rho_1)\,.
\nonumber
\qqq

\subsection{Multiplicative ergodic theorem and multiplicative large deviations}

\noindent The main general result about dynamical systems, the multiplicative
ergodic theorem of Oseledec \cite{Osel}, see also \cite{Ruelle}, states that 
under mild assumptions, the limits
\qq
\lim\limits_{t\to\pm\infty}\ {1\over t}\,\ln(W^TW)^\omega_t(\bm r)\ =:\ 
\Lambda_\pm^\omega(\bm r)
\nonumber
\qqq
exist for $M$-almost all $\,(\bm r,\omega)$. \,Besides,
\qq
\Lambda_\pm^\omega(\bm r)\ =\ O_\pm^\omega(\bm r)\,\,{\rm diag}[\lambda_1,
\cdots,\lambda_d]\m\,O_\pm^\omega(\bm r)^T
\label{lamb}
\qqq
where $\,\lambda_1\geq\cdots\geq\lambda_d\,$ are the Lyapunov exponents
that, due to the ergodicity assumption, are $\,(\bm r,\omega)$-independent. 
In particular, $\,\lambda_i\,$ are the limits when $\,t\to\pm\infty\,$
of the ratios $\,{\rho_i^\omega}_{\hspace{-0.05cm}t}(\bm r)/t\,$ 
for $\,M$-almost all $\,(\bm r,\omega)$.
\vskip 0.3cm

When all Lyapunov exponents are different, one may expect that for large but
finite time the distribution of the stretching exponents takes for 
$\,\rho_1>\cdots>\rho_d\,$ the large deviation form \cite{BF,CIRM}:
\qq
P_t(\bm\rho)\ \propto\ \ee^{-t\,H(\rho_1/t,\dots,\rho_d/t)}
\label{Pt}
\qqq
with a convex rate function $\,H\,$ attaining its minimal value equal to zero 
at the vector $\,\bm\lambda\,$ of the Lyapunov exponents. In particular, if 
$\,H\,$ is regular around $\,\bm\lambda\,$ one would obtain, as a corollary,
the multiplicative central limit result stating that 
$\,{1\over\sqrt{t}}(\bm\rho_t
-\bm\lambda t)\,$ tends when $\,t\to\infty\,$ to the vector of normal
variables with the inverse covariance given by the second derivative matrix 
$\,H''(\bm\lambda)$.  
\,To our knowledge, no general theorems assure existence of the multiplicative
large deviation regime, see \cite{BaxStr} for some partial results. 
There are, nevertheless, examples of random (and non-random \cite{Grass}) 
dynamical systems where the relation (\ref{Pt}) indeed holds. As has 
been already mentioned in Introduction, the examination of one of such 
examples is the main topic of the present paper. 
\vskip 0.3cm

For time reversible velocity ensembles, i.e. when the velocities 
$\,\bm u^\omega_t(\bm r)\,$ and $\,-\bm u^\omega_{-t}(\bm r)\,$ have the same 
distribution, the rate function $\,H\,$ possesses 
the Evans-Cohen-Morriss \cite{ECM} or Gallavotti-Cohen \cite{GC} 
type symmetry:
\qq
H(-\rho_d/t,\dots,-\rho_1/t)\ =\ H(\rho_1/t,\dots,\rho_d/t)\,
-\,\sum\limits_{i=1}^d\rho_i/t\,,
\label{GC}
\qqq
see \cite{BFF,FGV}. The relations of this type were discussed in a
closely related setup in \cite{BGG}.

\subsection{Stretching along the unstable flags}
\label{sec:sauf}

\noindent We shall call a family $\,F=(E_i)_{i=1}^d\,$ 
of $\,i$-dimensional subspaces 
\qq
\{0\}\,\subset\,E_{1}\,\subset\,\cdots\,\subset\,
E_{d-1}\,\subset\,E_{d}=\NR^d
\nonumber
\qqq
a \,{\bf flag}. \,An example is provided by the flag $\,F^0\,$ composed 
by the subspaces 
$\,E^0_i\,$ spanned by the first $\,i\,$ vectors of the canonical
basis of $\,\NR^d$. \,Clearly, the group $\,GL(d)\,$ acts on the space 
of flags. The subgroup $\,B\subset GL(d)\,$ composed of the upper 
triangular matrices preserves the flag $\,F_0\,$ and the Grassmannian  
$\,\Gr\,$ of all flags may be identified with the 
homogeneous space $\,GL(d)/B=O(d)/D\,$ where $\,D=O(d)\cap B\,$ is
the subgroup of the diagonal matrices with entries $\,\pm1$. \,We shall 
denote by $\,dF\,$ the normalized $\,O(d)$-invariant measure of $\,\Gr$.
\vskip 0.2cm

The flow $\,\Phi^\omega_t\,$ on the volume $\,V\,$ induces the
flow $\,\Psi^\omega_t\,$ on the product space $\,V\times\Gr\,$
defined by 
\qq
\Psi^\omega_t(\bm r,F)\ =\ (\Phi^\omega_t(\bm r),\m W^\omega_t(\bm r)F)\,.
\nonumber
\qqq
Mimicking the constructions from Sect.\,\ref{sec:nim} of the natural
invariant measure, one may define the measures $\,d\sigma^\omega(d\bm r,dF)\,$ 
on $\,V\times\Gr\,$ by the relation
\qq
\lim\limits_{s\to-\infty}\,\,{1\over|V|}\int\limits_{V\times\Gr} 
f(\Psi^{\omega_{s}}_{-s}(\bm r,F))\m\,d\bm r\,dF\ =:\ \int\limits_{V\times\Gr} 
f(\bm r,F)\m\,\sigma^\omega(d\bm r,dF)
\nonumber
\qqq
if the limit exists for continuous functions $\,f\,$ for almost all 
$\,\omega$. \,Clearly, the measures $\,\sigma^\omega\,$ depend
only on the past velocities and the formula
\qq
\int\limits_{V\times\Gr} f(\Psi^\omega_t(\bm r,F))\m\,
\sigma^\omega(d\bm r,dF)\ =\ 
\int\limits_{V\times\Gr}f(\bm r,F)\,\sigma^{\omega_{t}}(d\bm r,dF)\,.
\label{mdys}
\qqq
analogous to Eq.\,(\ref{mdy}) holds. Out of the measures $\,\sigma^\omega$, 
\,one may synthesize a measure $\,\Sigma(d\bm r,dF,d\omega)\,$ on the product 
space $\,V\times\Gr\times\Omega\,$ by the relation
\qq
\int\limits_{V\times\Gr\times\Omega}f(\bm r,F,\omega)\,
\Sigma(d\bm r,dF,d\omega)\ =\ 
\bigg\langle\int\limits_{V\times\Gr}f(\bm r,F,\omega)\,\sigma^\omega(d\bm r,dF)
\bigg\rangle.
\nonumber
\qqq
\vskip 0.2cm

For $\,F=OF^0\,$ with $\,O\in O(d)$, consider the Iwasawa decomposition 
of the invertible matrix $\,\tilde W=W^\omega_t(\bm r)\,O\m$:
\qq
\tilde W\ =\ O'\,{\rm diag}[\ee^{\eta_1},\dots,\ee^{\eta_d}]\,N
\nonumber
\qqq
with $\,O'\,$ orthogonal and $\,N\,$ upper-triangular with units on 
the diagonal. 
%The relations (\ref{WEE}) imply that
%$\,O'=\ O_-^{\omega_{_t}}(\Phi_t^\omega(\bm r))\,$ for an appropriate
%choice of the latter matrix. 
The exponents $\,\eta_i={\eta_i^\omega}_{\hspace{-0.05cm}t}(\bm r,F)\,$ 
do not depend on the freedom in the choice of $\,O\,$. \,We shall call 
them the \,{\bf stretching exponents along the flag} $\,F$. \,With 
$\,(\bm r,F,\omega)\,$ sampled w.r.t. the probability measure 
$\,\Sigma(d\bm r,dF,d\omega)$, they become
random variables. Their joint time $\,t\,$ PDF will be denoted 
$\,Q_t(\bm\eta)$.
\vskip 0.2cm

When all the Lyapunov exponents are different then the orthogonal
matrices $\,O_\pm^\omega(\bm r)\,$ in Eq.\,(\ref{lamb}) are determined
modulo the right multiplication by matrices from $\,D$. \,In particular, 
$\,O_-^\omega(\bm r)\,$ defines a flag $\,F^\omega(\bm r)\,$
of subspaces $\,E_i^\omega(\bm r)=O_-^\omega(\bm r)E^0_i\,$ of 
(less and less) unstable directions of the flow. In this case,
\qq
\int\limits_{V\times\Gr}f(\bm r,F)\m\,\sigma^\omega(d\bm r,dF)\ 
=\ \int\limits_V f(\bm r,F^\omega(\bm r))\m\,\mu^\omega(d\bm r)\,,
\nonumber
\qqq
i.e.\, the measure $\,\sigma^\omega(d\bm r,dF)\,$ is concentrated
in the direction of the Grassmannian $\,\Gr\,$ on the unstable flags.
The subspaces $\,E_{i}^\omega(\bm r)\,$ may be characterized by the property
that for $\,0\not=e\in E_{i}^\omega(\bm r)\setminus 
E_{i-1}^\omega(\bm r)$,
\qq 
\lim\limits_{t\to-\infty}\,\,{1\over t}\,\ln\Vert W^\omega_t(\bm r)e\Vert\,
=\,\lambda_i
\nonumber
\qqq
with the limit testing the far past asymptotics.
The unstable flags $\,F^\omega(\bm r)\,$ depend only on the velocities
at the negative times and are covariant under the 1-parameter 
dynamics (\ref{1pg}):
\qq
W^\omega_t(\bm r)\,F^\omega(\bm r)\ 
=\ F^{\omega_t}(\Phi_t^\omega(\bm r))\,.
\nonumber
\qqq
We shall call the exponents $\,{\eta_i^\omega}_{\hspace{-0.05cm}t}(\bm r)=
{\eta_i^\omega}_{\hspace{-0.05cm}t}(\bm r,F^\omega(\bm r))\,$ 
the \,{\bf stretching exponents along the unstable flags}. 
\vskip 0.2cm

The exponents $\,{\eta_i^\omega}_{\hspace{-0.05cm}t}(\bm r)\,$ are not 
equal to the stretching exponents $\,{\rho_i^\omega}_{\hspace{-0.05cm}t}
(\bm r)\,$ introduced previously. \,In particular, they are not necessarily 
non-increasing with $\,i\,$ although again the ratios 
$\,{\eta_i^\omega}_{\hspace{-0.05cm}t}(\bm r)/t\,$ tend
for $\,M$-almost all $\,(\bm r,\omega)\,$ to the ordered Lyapunov exponents 
$\,\lambda_i\,$ when $\,t\to\pm\infty$. \,Although the PDFs 
$\,Q_t(\bm\eta)\,$ and $\,P_t(\bm\rho)\,$ are, in general, different 
(in particular, the latter is non-zero only for ordered arguments), 
we shall see below that the large deviation parts of $\,P_t\,$ and $\,Q_t\,$ 
are closely related. In fact, the stretching exponents $\,\eta_i\,$ are 
more natural objects than the exponents $\,\rho_i\,$ and, as we shall see 
in examples, their evolution is often simpler to describe.

\nsection{Lagrangian flow in the Kraichnan model}
\subsection{Kraichnan ensemble of velocities}

\noindent The Kraichnan ensemble of $\,d$-dimensional velocities 
$\,{\bm u}^\omega(t,{\bm r})\,$ is the Gaussian random ensemble 
characterized by vanishing mean and time decorrelated 
covariance:
\qq
\Big\langle\,{u^\omega_t}^{i}({\bm r})
\,{u^\omega_{t'}}^{j}({\bm r'})\,
\Big\rangle\ =\ \delta(t-t')\,D^{ij}({\bm r},{\bm r}')\,.
\nonumber
\qqq
In particular, we shall consider a homogeneous ensemble in the $d$-dimensional 
periodic box $\,V\,$ of side $\,L\,$ with the spatial covariance given 
by the Fourier series:
\qq
D^{ij}({\bm r},{\bm r}')\ =\ \sum\limits_{{\bm k}\in{2\pi\over L}{\bf Z}^d} 
[(1-\wp)\delta^{ij}-(1-\wp d)k^ik^j]\,e^{i{\bm k}
\cdot({\bm r}-{\bm r'})}\,\hat d(|{\bm k}|)\,,
\label{Dij}
\qqq
as often used in numerical simulations. The spectral function 
$\,\hat d\,$ will be assumed fast decreasing or 
of compact support so that the resulting covariance and, consequently,
almost all velocity realizations $\,\bm u^\omega_t(\bm r)\,$ are 
smooth in space. 
The simplest would be to take $\,\hat d\,$ supported only on the modes 
with $\,|{\bm k}|={2\pi\over L}$, \ i.e. on the lowest nontrivial ones. 
The parameter $\,\wp\,$ in Eq.\,(\ref{Dij}) is called the {\bf\,compressibility 
degree}. \,It is equal to the ratio of the covariances 
$\,\langle(\sum\nabla_i{u^\omega}^i)^2\rangle\Big/\langle\sum(\nabla_i{u^\omega}^j)^2
\rangle\,$ and is contained between zero and 
one. Vanishing $\,\wp\,$ corresponds to an incompressible flow and 
$\,\wp=1\,$ to a gradient one. 
\vskip 0.2cm

The Lagrangian flow in the Kraichnan ensemble of velocities is 
defined by the ordinary differential equation (\ref{Lf}).
Because of the white-noise temporal behavior of the Kraichnan velocities, 
Eq.\,(\ref{Lf}) becomes, however, a stochastic differential equation and,
in line with more standard notations, will be written in the form
\qq
d\bm R\ =\ \bm u^\omega_t(\bm r)\,dt\,.
\nonumber
\qqq
In principle, it requires a choice of the stochastic convention, like It\^o's 
or Stratonovich's one, but in the case in question both choices 
lead to the same 
solutions (this is due to the vanishing of $\,\nabla_{r^j}D^{ij}({\bm 0},
{\bm 0})$). \,As before, one obtains from the solutions a family 
$\,\Phi_\omega^t\,$ of smooth random maps \cite{Kuni}.
\vskip 0.3cm

For the Kraichnan model, the convergence (\ref{mu}) takes
place in the $\,L^2\,$ norm of the Gaussian process. Besides, 
due to the homogeneity of the velocity ensemble,
\qq
\int\limits_{V\times\Omega}f(\bm r)\,M(d\bm r,d\omega)\ =\ 
\bigg\langle\int\limits_Vf(\bm r)\,\mu^\omega(d\bm r)\bigg\rangle\ =
\ {1\over|V|}\int\limits_V f(\bm r)\,d\bm r\,.
\label{sm1}
\qqq
Similarly, due to the homogeneity, 
\qq 
\int\limits_{V\times\Gr\times\Omega}f(\bm r,F)\,\m\,\Sigma(d\bm r,dF,d\omega)
\,=\,\bigg\langle\int\limits_{V\times\Gr}f(\bm r,F)\m\,\sigma^\omega
(d\bm r,dF)\bigg\rangle\,=\,{1\over|V|}\int\limits_{V\times\Gr}
f(\bm r,F)\m\,d\bm r\,\chi(dF)\quad
\label{sm2}
\qqq
for some probability measure $\,\chi(dF)\,$ on $\,\Gr$. \,Note that 
averaging Eq.\,(\ref{mdys}) with respect to the probability measure 
$\,P(d\omega)\,$ for functions $\,f\,$ independent of $\,\bm r$, 
\,one infers that
\qq
\int\limits_{\Gr}\Big\langle f(W^\omega_t(\bm r_0)F)\Big\rangle\m\,\chi(dF)
\ =\ {1\over|V|}\int\limits_{V\times\Gr}\Big\langle f(W^\omega_t(\bm r)F)
\Big\rangle\m\,d\bm r\,\chi(dF)\ =\ \int\limits_{\Gr}f(F)\m\,\chi(dF)\,,
\label{mdyc}
\qqq
i.e. that the measure $\,\chi(dF)\,$ is invariant under the process 
$\,W^\omega_t$.

\subsection{Tangent process in Kraichnan velocities}

\noindent Further simplifications appear in the statistics of the 
tangent process $\,W^\omega_t(\bm r)$. \,For positive $\,t$, 
$\,W^\omega_t(\bm r)\,$ depends 
only on the velocities at positive times and $\,\mu^\omega(\bm r)\,$ on 
the velocities at negative times. The temporal decorrelation of the 
Kraichnan velocities implies then for any function $\,f\,$ of invertible 
$\,d\times d\,$ matrices and for $\,t\geq0\,$ the factorization
\qq
&\displaystyle{\int\limits_{V\times\Omega}f(W^\omega_t(\bm r))\m\,M(d\bm r,
d\omega)\ \equiv\ \bigg\langle\int\limits_V f(W^\omega_t(\bm r))
\m\,\mu^\omega(d\bm r)\bigg\rangle}&\cr 
&\displaystyle{=\ \bigg\langle\int\limits_V\Big\langle f(W^\omega_t(\bm r))
\Big\rangle\,\mu^\omega(d\bm r)\bigg\rangle\ 
=\ {1\over|V|}\int\limits_V\Big\langle f(W^\omega_t
(\bm r))\Big\rangle\,d\bm r\ =\ \Big\langle f(W^\omega_t(\bm r_0))
\Big\rangle}&
\nonumber
\qqq
where the last but one equality follows from the relation (\ref{sm1})
and the last one is again due to the homogeneity of the velocity ensemble.
We infer that it is enough to know the distribution of 
$\,W^\omega_t(\bm r_0)\,$ for one fixed $\,\bm r_0$. \,This
simplifies considerably the analysis of the statistics of the
stretching exponents in the Kraichnan model. 
\vskip 0.3cm

Let us suppose now that $\,f\,$ is a function on $\,GL(d)\,$
invariant under the right multiplication of its argument by
diagonal matrices with entries $\,\pm1\,$ so that 
$\,f(\tilde W)\,$ for $\,\tilde W=W^\omega_t(\bm r)\,O\,$ depends 
on $\,O\,$ only via the flag $\,F=OF^0\,$ (but is not, in general, 
a function of $\,W^\omega_t(\bm r)F\,$ only), see Sect.\,\ref{sec:sauf}.
Again due to the decorrelation of velocities at positive and negative 
times and the relation (\ref{sm2}),
\qq 
&\displaystyle{\int\limits_{V\times\Gr\times\Omega}f(\tilde W)
\,\,\Sigma(d\bm r,dF,d\omega)\ \equiv\ 
\bigg\langle\int\limits_{V\times\Gr} f(W^\omega_t(\bm r)\m O)
\m\,\sigma_\omega(d\bm r,dF)\bigg\rangle}&\cr
&\displaystyle{=\ {1\over|V|}\int\limits_{V\times\Gr}
\Big\langle f(W^\omega_t(\bm r)\,O)\Big\rangle\m\,d\bm r\,\chi(dF)
\ =\ \int\limits_{\Gr}\Big\langle f(W^\omega_t(\bm r_0)\,O)\Big\rangle
\m\,\chi(dF)}
\label{tWf}
\qqq
The last relation permits to simplify the analysis of the
statistics of the stretching exponents along the unstable flags.

\subsection{Multiplicative stochastic equation for the tangent process}

\noindent For fixed $\,\bm r_0$, \,the distribution of the  
tangent process $\,W^\omega_t(\bm r_0)\,$  may be obtained by solving 
the multiplicative stochastic equation
\qq
dW\ =\ S^\omega_t\m W\m dt\,,
\label{ldes}
\qqq
the stochastic version of Eq.\,(\ref{lde}), with initial condition 
$\,W_0=\Id$. \,The further crucial simplification, due to the time 
decorrelation and spatial homogeneity of the Kraichnan velocity 
ensemble, is that in Eq.\,(\ref{ldes}) one may take $\,(S^\omega_t)^i_{\,\,j}
=\nabla_j{u^\omega_t}^{i}(\bm r_0)\,$ instead 
of $\,(S^\omega_t)^i_{\,\,j}=\nabla_j{u^\omega_t}^{i}
(\bm R^\omega_t(r_0))$. \,In other words, the dependence on the trajectory 
$\,R^\omega_t(\bm r_0)\,$ may be dropped from $\,S^\omega_t\,$ as long 
as we are interested in the distribution of $\,W^\omega_t(\bm r_0)\,$ 
for fixed $\,\bm r_0$, \,provided we consider 
the differential equation (\ref{lde}) with the It\^o convention (here 
the convention does matter, see Appendix A in \cite{FGV}). 
\vskip 0.3cm

Summarizing, the strain process $\,S^\omega_t\,$ in Eq.\,(\ref{ldes}) may be 
taken as the matrix-valued white noise with mean zero and the covariance
\qq
\Big\langle (S^\omega_t)^i_{\,\,k}\,\,(S^\omega_{t'})^j_{\,\,l}\Big\rangle
\ =\ \delta(t-t')\,\nabla_{r^k}\nabla_{r'^l}D^{ij}(\bm 0,\bm 0)\ 
=:\ \delta(t-t')\,C^{ij}_{kl}\,.
\label{SS}
\qqq
For the spatial velocity covariance $\,D^{ij}(\bm r,\bm r')\,$ given by the
Fourier series (\ref{Dij}),
\qq
C^{ij}_{kl}\ &=&\sum\limits_{\vec{k}\in{2\pi\over L}{\bf Z}^d}
\Big[(1-\wp)\delta^{ij}-(1-\wp d)k^ik^j\Big]k_kk_l\,\hat d(|\vec{k}|)\cr 
&=&2\alpha\,\delta^{ij}_{kl}\,+\,\beta\,(\delta^i_k\delta^j_l+\delta^i_l\delta^j_k)\,
+\,\gamma\,\delta^{ij}\delta_{kl}
\label{C4}
\qqq
where $\,k^i\equiv k_i\,$ and $\,\delta^{ij}_{kl}\,$ is equal to 1 if $\,i=j=k=l\,$ 
and to zero otherwise. The compressibility degree
\qq
\wp\ =\ {\Big\langle(\tr\,S^\omega)^2\Big\rangle\over\Big\langle\tr\,
{S^\omega}^{\hspace{-0.03cm}T}S^\omega\Big\rangle}\ 
=\ {2\alpha+(d+1)\beta+\gamma\over2\alpha+2\beta+d\gamma}\,.
\nonumber
\qqq
The positivity of the covariance imposes the inequalities
\qq
\quad\gamma\geq|\beta|\,,\quad 4\alpha+(d+2)\beta+2\gamma\geq d|\beta|\,.
\nonumber
\qqq
We shall exclude the trivial case $\,\alpha=\beta=\gamma=0$.
\,The case of vanishing $\,\alpha\,$ corresponds to the isotropic situation
when the distributions of $\,S^\omega\,$ and of $\,OS^\omega O^T\,$ 
coincide for any orthogonal matrix $\,O$. \,The $\alpha$-term breaks 
the $\,O(d)$-invariance of the distribution of $\,S^\omega\,$ to the 
discrete subgroup of the symmetries of a cube. It is the source of a 
small scale anisotropy that occurs generically in the Kraichnan flow in 
a periodic box. Indeed, vanishing of $\,\alpha\,$ requires a fine tuning 
of the spectral density $\,\hat d(|\bm k|)$. For example, when $\,\hat d\,$ 
is non-zero only for $\,|\bm k|={2\pi\over L}\,$ then necessarily 
$\,\alpha\not=0$.

\subsection{Generator of the tangent process}

\noindent The generator of the process $\,W_t$ satisfying stochastic
equation Eq.\,\,(\ref{ldes}), \,i.e.\,\,the operator $\,\CL\,$ such that
for any regular function $\,f\,$ on the group $\,GL(d)$,
$${d\over dt}\Big\langle f(W)\Big\rangle\ 
=\ \Big\langle(\CL f)(W)\Big\rangle\,,$$
is given by the formula
\qq
\CL\,&=&\,{_1\over^2}
\sum\limits_{i,j,k,l,\atop n,m=1}^d\Big(2\alpha\,\delta^{ij}_{kl}\,
+\,\beta\,(\delta^i_{\,\,k}\delta^j_{\,\,l}\,+
\,\delta^i_{\,\,l}\delta^j_{\,\,k})\,+\,\gamma\,
\delta^{ij}\delta_{kl}\Big)\,W^k_{\,\,m}W^l_{\,\,n}
\,\partial_{W^i_{\,\,m}}\partial_{W^j_{\,\,n}}\cr
&=&\,\alpha\sum\limits_{i=1}^d(\CE_i^{\,\,i})^2\,+\,
{_1\over^2}(\beta+\gamma)\CE^2-{_1\over^2}\gamma\CJ^2+{_1\over^2}\beta\m\CD^2
-(\alpha+{_1\over^2}(d+1)\beta+{_1\over^2}\gamma)\CD\,,
\label{cl}
\qqq
where, for $\,E^i_{\,\,j}\,$ denoting the basic matrices with the matrix
elements $\,(E^i_{\,\,j})^k_{\,\,l}=\delta^{ik}\delta_{jl}$,
\qq
&\displaystyle{(\CE_i^{\,\,j}f)(W)={_d\over^{ds}}|_{_{s=0}}
f(\ee^{-s\,E^i_{\,\,j}}W)=-\sum\limits_k W^j_{\,\,k}
\partial_{W^i_{\,\,k}}\,f(W)\,,\quad
\CE^2=\sum\limits_{i,j}\CE_i^{\,\,j}\CE_j^{\,\,i}\,,}&\cr
&\displaystyle{\CJ^2=-{_1\over^2}\sum\limits_{i,j}(\CE_i^{\,\,j}
-\CE_j^{\,\,i})^2\,,\qquad\CD\ =\ {_d\over^{ds}}|_{_{s=0}}f(\ee^sW)
=-\sum\limits_i\CE_i^{\,\,i}\,.}&
\nonumber
\qqq
Note that $\,\CE^2\,$ is the quadratic Casimir of 
$\,gl(d)$, $\,\CJ^2\,$ the one
of $\,so(d)\,$ and $\,\CD\,$ the generator of the dilations. 
Formula (\ref{cl}) goes back to ref.\,\cite{SS} where it was
discussed for the isotropic incompressible case.
For any values of the parameters, the generator $\,\CL\,$
commutes with the right regular action 
\qq
(R_Mf)(W)\,=\,f(WM)
\nonumber
\qqq
of $\,GL(d)\,$ on functions on itself.

\subsection{Stretching exponents in the isotropic case}

\noindent The isotropic case with $\,\alpha=0\,$ has been treated in
ref.\,\cite{BF,BFF}, see also \cite{SS,SM}.\,Here $\,\CL\,$ commutes also
with the left action of $\,O(d)\,$ given by
\qq
(L_Of)(W)\,=\,f(O^{-1}W)\,.
\nonumber
\qqq
In particular, $\,\CL\,$ preserves the space of functions invariant 
under the left and right action of $\,O(d)$, \,i.e. functions 
$\,f(\bm\rho)\,$ that depend on $\,W\,$ only through the stretching 
exponents, see Eq.\,(\ref{stre}). In other words, the stretching
exponents evolve independently of the angles of the $\,O(d)\,$
matrices in the decomposition (\ref{stre}). On functions $\,f(\bm\rho)$,
the generator $\,\CL\,$ reduces to the operator 
\qq
\CL_{\bm\rho}\ =\ {_{\beta+\gamma}\over^2}\Big(\sum\limits_{i=1}^d{\partial^2
\over\partial\rho_i^2}
\,+\,\sum\limits_{i\not=j}\coth(\rho_i-\rho_j)\,{\partial\over\partial\rho_i}\Big)\,
+\,{_\beta\over^2}\Big(\sum\limits_{i=1}^d{\partial\over\partial\rho_i}\Big)^2\,-\,
{_{(d+1)\beta+\gamma}\over^2}\sum\limits_{i=1}^d{\partial\over\partial
\rho_i}\,.
\nonumber
\qqq
The right hand side is the generator of the diffusion process $\,\bm\rho_t\,$
satisfying the stochastic differential equations \cite{BF}
\qq
d\rho_i\ =\ {_{\beta+\gamma}\over 2}\sum\limits_{j\not=i}\coth(\rho_i-\rho_j)\m dt\,-\,
{_{(d+1)\beta+\gamma}\over^2}dt\,+\,{\zeta_i}_t\m dt
\label{ster}
\qqq
where $\,\bm\zeta_t\,$ is the white noise with the covariance
\qq
\Big\langle{\zeta_i}_t\,{\zeta_j}_{t'}\Big\rangle\ 
=\ [(\beta+\gamma)\delta_{ij}\,+\,\beta]\,\delta(t-t')\,.
\nonumber
\qqq
The time $\,t\,$ PDF $\,P_t(\bm\rho)\,$ may be expressed in terms of the heat
kernel of the Calogero-Sutherland Hamiltonian \cite{SM,CIRM}.
The Lyapunov exponents are given by \cite{LJ,Bax}
\qq
\lambda_i\ =\ {_{\beta+\gamma}\over^2}(d-2i+1)-{_{(d+1)\beta+\gamma}\over^2}
\nonumber
\qqq
and are all different. The large deviation form of $\,P_t(\bm\rho)\,$
is easy to obtain by the following heuristic considerations \cite{BF}.
Since $\,|\rho_i-\rho_j|\,$ for $\,i\not=j\,$ grows approximately 
linearly with time, at long times $\,\coth(\rho_i-\rho_j)\approx\pm1\,$ and 
the operator $\,\CL_{\bm\rho}\,$ should reduce to the asymptotic form
\qq
\CL^{as}_{\bm\rho}\ =\ {_{\beta+\gamma}\over^2}\sum\limits_{i=1}^d{\partial^2
\over\partial\rho_i^2}\,+\,{_\beta\over^2}\Big(\sum\limits_{i=1}^d{\partial
\over\partial\rho_i}\Big)^2\,+\,\sum\limits_{i=1}^d\lambda_i
{\partial\over\partial\rho_i}\,.
\nonumber
\qqq
Similarly, the stochastic equation (\ref{ster}) simplifies at long times to
\qq
d\rho_i\ =\ \lambda_i\m dt\,+\,{\zeta_i}_t\m dt
\nonumber
\qqq
The long-time asymptotics of $\,\rho_i\,$ is now easy to find leading 
to the large deviation form (\ref{Pt}) of the PDF 
$\,P_t(\bm\rho)\,$ with the quadratic rate function \cite{BF,BFF}
\qq
H(\bm\rho/t)\ =\ {_1\over^{2(\beta+\gamma)}}\Big[\sum\limits_{i=1}^d
({_{\rho_i}\over^t}-\lambda_i)^2\,-\,{_\beta\over^{(d+1)
\beta+\gamma}}\Big(\sum\limits_{i=1}^d({_{\rho_i}\over^t}
-\lambda_i)\Big)^{\hspace{-0.07cm}2}\Big]
\label{LDH}
\qqq
taking its minimal value at $\,\bm\rho/t=\bm\lambda\,$ and possessing
the symmetry (\ref{GC}).
\vskip 0.3cm

Let us turn now to the stretching exponents $\,{\bm\eta}_t^\omega(\bm r,F)\,$
along flags $\,F=OF^0$. \,Since $\,\eta_i\,$ are defined by the
Iwasawa decomposition of the matrix $\,\tilde W=
W^\omega_t(\bm r)\,O$, \,functions of $\,\bm\eta\,$
may be identified with functions $\,f\,$ of $\,\tilde W\,$ invariant
under the right action of upper-triangular matrices $\,N\,$ with
units on the diagonal and the left action by orthogonal matrices 
$\,O\,$ (such functions are necessarily 
invariant under the right action by diagonal matrices with entries 
$\,\pm1$). The calculation of the average of $\,f(\tilde W))\,$ with 
respect to the measure $\,\Sigma(d\bm r,dF,d\omega)\,$ is now simplified 
by Eq.\,(\ref{tWf}). For fixed $\,\bm r_0\,$ and $\,O\in O(d)$, \,the 
statistics of $\,W^\omega_t(\bm r_0)\,O\,$ may be found by solving 
the It\^o stochastic equation (\ref{ldes}) with the initial condition 
$\,W_0=O$. \,It follows that
\qq
{d\over dt}\Big\langle f(W^\omega_t(\bm r_0)\,O)\Big\rangle
\ =\ \Big\langle(\CL f)(W^\omega_t(\bm r_0)\,O)\Big\rangle\,.
\nonumber
\qqq
In the isotropic case with $\,\alpha=0$, the generator \,$\CL\,$ preserves 
the space of function invariant under the right action by upper-triangular
matrices $\,N\,$ with units on the diagonal and under the left action 
of orthogonal matrices $\,O\,$ and reduces on such functions to the operator
\qq
\CL_{\bm\eta}\ \  =\ {_{\beta+\gamma}\over^2}\sum\limits_{i=1}^d{\partial^2
\over\partial\eta_i^2}\,+\,{_\beta\over^2}\Big(\sum\limits_{i=1}^d{\partial
\over\partial\eta_i}\Big)^2\,+\,\sum\limits_{i=1}^d
\lambda_i{\partial\over\partial\eta_i}
\nonumber
\qqq
of the form coinciding for all times with the asymptotic form 
$\,\CL^{as}_{\bm\rho}\,$ of $\,\CL_{\bm\rho}$. \,As the result,
\qq
\Big\langle f(W^\omega_t(\bm r_0)\,O)\Big\rangle\ =\ \int
\ee^{-t\CL_{\bm\eta}}\hspace{-0.05cm}(\bm0,\bm\eta)\m\,f(\bm\eta)\m\,d\bm\eta
\ =\ {\int f(\bm\eta)\m\,\ee^{-t\,H(\eta^1/t,\dots,\eta^d/t)}
\m\,d\bm\eta\over\int\ee^{-t\,H(\eta^1/t,\dots,
\eta^d/t)}\m\,d\bm\eta}
\nonumber
\qqq
with $\,H\,$ given by Eq.\,(\ref{LDH}). In particular, the $\,O$-dependence
drops out and the integral over the Grassmannian $\,\Gr\,$ on the right hand 
side of Eq.\,(\ref{tWf}) becomes trivial (in fact, in the isotropic case,
$\,\chi(dF)=dF\,$ i.e. it is $\,O(d)$-invariant measure on $\,\Gr$). 
\,We infer that the time $\,t\,$ PDF $\,Q_t(\bm\eta)\,$  of the stretching 
exponents along the (unstable) flags is Gaussian for all times:
\qq
Q_t(\bm\eta)\ =\ {\ee^{-t\,H(\eta^1/t,\dots,\eta^d/t)}\over
\int\ee^{-t\,H(\eta^1/t,\dots,\eta^d/t)}\m\,d\bm\eta}.
\nonumber
\qqq
In particular, its large deviation form coincides with that for the stretching 
exponents $\,\rho_i\,$ except that in the latter case, it is restricted to the 
region where $\,\rho_1>\cdots>\rho_d$. \,We shall see below that such 
coincidence of the large deviation statistics for $\,\bm\rho\,$ and 
$\,\bm\eta\,$ holds in more general situations.

\nsection{Kraichnan flow in a periodic square}
\subsection{Generator of the $2d$ tangent process}

\noindent We shall discuss here the statistics of the solutions of the
multiplicative stochastic equation (\ref{ldes}) for $\,2\times2\,$ matrices
with the covariance of the white noise strain $\,S^\omega_t\,$ given by
Eqs.\,(\ref{SS}) and (\ref{C4}) with $\,d=2$. \,The overall time 
scale of the strain is set by
\qq
{{4\m\delta(0)}\over{\langle\tr\,{S^\omega}^{\hspace{-0.02cm}T}S^\omega
\rangle}}\ =\ {1\over{\alpha+\beta+\gamma}}\ \equiv\ \tau\,.
\nonumber
\qqq
The distribution of $\,S^\omega_t\,$ is invariant under the $\,90^o\,$
rotations and reflections in the coordinate axis. If $\,O_{\pi\over4}\,$ is 
the rotation by $\,45^o$, \,i.e.
\qq
O_{\pi\over4}\ =\ {_1\over^{\sqrt{2}}}\left(\matrix{1&1\cr-1&1}\right)
\nonumber
\qqq
then $\,{S'_t}^\omega=O_{\pi\over4}S^\omega_tO_{\pi\over4}^T\,$ has the 
similar distribution to that of $\,S^\omega_t\,$ but with parameters 
\qq
\alpha'=-\alpha\,,\quad\ \beta'=\alpha+\beta\,,\quad\ \gamma'=\alpha+\gamma
\nonumber
\qqq
Note that $\tau'=\tau$, $\,2\gamma'+\alpha'=2\gamma+\alpha\,$ and that 
the compressibility degree is the same for both sets 
of the parameters. We shall call the ratio
\qq
\kappa\ =\ {|\alpha|\tau}\ =\ {|\alpha'|\tau'}
\nonumber
\qqq
the {\bf\,anisotropy degree}. \,It measures the difference between the 
covariances 
of the the processes $\,S^\omega_t\,$ and $\,{S'_t}^\omega\,$ relative to 
$\,\Big\langle\tr\,{S^\omega}^{\hspace{-0.02cm}T}S^\omega\Big\rangle\,$ and 
it is contained between zero and one.  We shall use the dependence
on $\,\kappa\,$ to measure the influence of the anisotropy on the 
distribution of the tangent process $\,W_t$. \,For $\,W_t\,$ solving 
Eq.\,(\ref{ldes}), the distribution of $\,W'_t=O_{\pi\over4}W_t
O_{\pi\over4}^T\,$ will coincide with that of the solution of 
Eq.\,(\ref{ldes}) for the primed values of the parameters. As the result, 
the distribution of the stretching exponents 
$\,{\bm\rho^\omega_t}(\bm r)\,$ as well as that of 
$\,{\bm\eta^\omega_t}(\bm r)\,$ for the two sets of parameters are 
identical. Below, we shall then restrict ourselves to the case 
with $\,\alpha\geq0$.
\vskip 0.3cm

The parametrization (\ref{stre}) takes in two dimensions the form:
\qq
W\ =\ \left(\matrix{\cos{\phi\over 2}&\sin{\phi\over 2}\cr
-\sin{\phi\over 2}&\cos{\phi\over 2}}\right)
\left(\matrix{\ee^{\m\rho_1}&0\cr0&\ee^{\m\rho_2}}\right)
\left(\matrix{\cos\psi&\sin\psi\cr-\sin\psi&\cos\psi}\right)
\nonumber
\qqq
(we may assume that $\,\det\,W=1$). \,Hence functions of $\,W\,$ may 
be viewed as functions of two angles and the stretching exponents 
and  satisfying the relations 
\qq
f(\phi,\rho_1,\rho_2,\psi)\,&=&\,f(\phi+\pi,\rho_2,\rho_1,
\psi-{_\pi\over^2})\,,\label{per1}\cr
\qquad f(\phi,\rho_1,\rho_2,\psi)\,&=&\,f(\phi,\rho_1,\rho_2,\psi+2\pi)
\nonumber
\qqq
that permit to restrict the parameters to the region $\,\rho_1\geq\rho_2\,$ 
and $\,0\leq\phi,\psi\leq2\pi$. \,For the generator of the
tangent process given by Eq.\,(\ref{cl}) one obtains the following 
complicated expression:
\qq
\CL\ &=&\ {_1\over^2}\alpha\Big[\partial_{\rho_1}+\,\partial_{\rho_2}\Big]^2
\cr 
&&+\,{_1\over^2}\alpha\Big[2\,\sin{\phi}\,\coth(\rho_1-\rho_2)\,\partial_\phi
\,-\,\cos{\phi}\,(\partial_{\rho_1}-\partial_{\rho_2})\,
-\,\sin{\phi}\,\sinh^{-1}(\rho_1-\rho_2)\,\partial_\psi\Big]^2\cr
&&+\,{_1\over^2}(\beta+\gamma)\Big[\partial_{\rho_1}^{\,2}
+\,\partial_{\rho_2}^{\,2}
\,+\,2\,\sinh^{-2}(\rho_1-\rho_2)\,\partial_\phi^2\,+\,
{_1\over^2}\,\sinh^{-2}(\rho_1-\rho_2)\,\partial_\psi^2\,\cr
&&-\,2\,\cosh(\rho_1-\rho_2)\,\sinh^{-2}(\rho_1-\rho_2)\,
\partial_\phi\partial_\psi\,
+\,\coth(\rho_1-\rho_2)\,(\partial_{\rho_1}-\partial_{\rho_2})\Big]\cr
&&+\,2\,\gamma\,\partial_\phi^2\,+\,{_1\over^2}\beta\Big[
\partial_{\rho_1}+\,\partial_{\rho_2}\Big]^2\,-\,{_1\over^2}(2\alpha
+3\beta+\gamma)\,\Big[\partial_{\rho_1}+\,\partial_{\rho_2}\Big]\,.
\label{comex}
\qqq
$\CL\,$ is self-adjoint in the $\,L^2\,$ scalar product with the measure 
$\,\ee^{-\rho_1-\rho_2}\sinh|\rho_1-\rho_2|\m d\phi\m d\rho_1d\rho_2
d\psi$. \,Note that $\,\CL\,$ commutes separately with the translation of 
$\,\phi\,$ by $\,\pi$, with permutation of $\,\rho_i\,$ and with 
the arbitrary translations of $\,\psi$. \,Since at the end we shall 
be interested in the distribution of the (ordered) stretching exponents, 
we may right away restrict ourselves to the sector of functions that 
are periodic in $\phi\,$ of period $\pi$, \,even under the interchange 
of $\,\rho^i\,$ and independent of $\,\psi$. 
\,Upon setting
\qq
{_1\over^2}(\rho_1+\rho_2)\,\equiv\,r\,,\qquad \rho_1-\rho_2\,\equiv\,\rho
\nonumber
\qqq
$\CL\,$ reduces in the action on $\,\psi$-independent functions to
the operator $\,\CL_r+\CL_{\phi\rho}\,$ with
%\qq
%\CL_{\phi r\rho}\ &=&\ {_1\over^2}\alpha\,\partial_r^2\,+\,
%2\alpha\Big[\sin{\phi}\,\coth\rho\,\partial_\phi\,
%-\,\cos{\phi}\,\partial_\rho\Big]^2\cr
%&&\,+\,{_1\over^2}(\beta+\gamma)\Big[{_1\over^2}\partial_r^2\,+\,
%2\,\partial_\rho^2\,+\,2\,\sinh^{-2}\rho\,\partial_\phi^2\,+\,
%2\,\coth\rho\,\partial_\rho\Big]\cr
%&&\,+\,2\gamma\,\partial_\phi^2\,+\,{_1\over^2}\beta\,\partial_r^2
%\,-\,{_1\over^2}(2\alpha+3\beta+\gamma)\,\partial_r\,.
%\label{CL}
%\qqq
\qq
\CL_r\,\,\,&=&\,{_1\over^4}(2\alpha+3\beta+\gamma)\,\partial_r^2\,
-\,{_1\over^2}(2\alpha+3\beta+\gamma)\,\partial_r\,,\label{clr}\\
\CL_{\phi\rho}\,&=&\,2\alpha\Big[\sin{\phi}
\,\coth\rho\,\partial_\phi\,-\,\cos{\phi}\,
\partial_\rho\Big]^2\,+\,2\gamma\,\partial_\phi^2\cr
&&\,+\,{_1\over^2}(\beta+\gamma)\Big[2\,\partial_\rho^2\,
+\,2\,\sinh^{-2}\rho\,\partial_\phi^2\,+\,
2\,\coth{\rho}\,\partial_\rho\Big]\,.
\label{phirho}
\qqq
It follows that at all times the processes $\,r_t\,$ and $\,(\rho_t,\phi_t)$, 
\,starting at $\,r_0=0=\rho_0\,$ and $\,\phi_0=0$, \,are independent. 

\subsection{Sum of Lyapunov exponents}

\noindent The PDF of $\,r_t\,$ takes for all times the Gaussian large 
deviation form since $\,\CL_r\,$ has constant coefficients:
\qq
P_t(r)\ =\ {1\over\sqrt{\pi(2\alpha+3\beta+\gamma)t}}\,\ee^{-t\m H_r(r/t)}
\label{rld}
\qqq
with the quadratic rate function
\qq
H_r(r/t)\ =\ {\left({r/t}+{1\over2}(2\alpha+3\beta+\gamma)\right)^2
\over2\alpha+3\beta+\gamma}\,.
\nonumber
\qqq
In particular, when $\,t\to\infty$, \, the PDF of $\,r\,$ concentrates
at $\,{r\over t}=-{1\over2}(2\alpha+3\beta+\gamma)\,$ which gives the half 
of the sum of the Lyapunov exponents:
\qq
\lambda_1+\lambda_2\ =\ -(2\alpha+3\beta+\gamma)\,.
\label{slexp}
\qqq
Note that if one normalizes the Lyapunov exponents by multiplying them by 
the overall time scale $\,\tau=(\alpha+\beta+\gamma)^{-1}\,$ then
one obtains the relation
\qq
(\lambda_1+\lambda_2)\tau
\ =\ -2\wp
\label{sumf}
\qqq
stating that the normalized sum of the Lyapunov exponents is directly
tied to the compressibility degree.
\vskip 0.3cm

The Evans-Cohen-Morriss-Gallavotti-Cohen symmetry (\ref{GC})
involves here only the large deviations of $\,r_t\,$ and reduces to 
the identity
\qq
H_r(-r/t)\ =\ H_r(r/t)\,-\,2r/t\,.
\nonumber
\qqq 

\subsection{Asymptotic form of the $\,(\rho,\phi)$-process}

\noindent It remains to find the large-time form of the joint PDF of 
$\,\phi\,$ and $\,\rho$. \,Unlike in the isotropic situation, the evolution of 
$\,\rho\,$ does not decouple from the angle $\,\phi\,$ and both have 
to be treated at the same time. Recall that we may restrict ourselves
to the sector of functions of period $\,\pi\,$ in $\,\phi\,$ and even
in $\,\rho\,$ so that we may consider only $\,\rho\geq0\,$ (i.e. $\,\rho_1
\geq\rho_2$) \,imposing the appropriate boundary condition at $\,\rho=0\,$
(recall that $\,\CL\,$ acts at smooth functions on $\,GL(d)$). \,At 
long times, $\,\rho/t\,$ will still concentrate at a single value equal 
to the difference of the Lyapunov exponents. Anticipating the latter to be 
strictly positive, we infer that at large $\,t\,$ the process $\,\rho_t\,$ 
will take predominantly large values $\,\propto t$. \,Consequently, we should 
be able to replace $\,\coth{\rho}\,$ by $\,1\,$ in the expression 
(\ref{phirho}) for the generator $\,\CL_{\phi\rho}\,$ and drop the term 
with $\,\sinh^{-2}{\rho}\,$ reducing $\,\CL_{\phi\rho}\,$ to the asymptotic 
form:
\qq
\CL^{as}_{\phi\rho}\ \,=\,\ 2\alpha\Big[\sin{\phi}
\,\partial_\phi\,-\,\cos{\phi}\,
\partial_\rho\Big]^2\,+\,2\gamma\,\partial_\phi^2
\,+\,(\beta+\gamma)\Big[\partial_\rho^2\,+\,
\partial_\rho\Big].
\label{gg}
\qqq
Note on the margin that under a similar 
transformation applied to operator $\,\CL\,$ of Eq.\,(\ref{comex}),
all terms involving the angle $\,\psi\,$ drop out implying that $\,\psi\,$
becomes frozen at long times, in agreement with the Oseledec theorem.
With an appropriate boundary conditions at $\,\rho=0$, 
\,the operator $\,\CL^{as}_{\phi\rho}\,$ is self-adjoint with respect to 
the $\,L^2\,$ scalar product with the measure 
$\,\ee^{\rho}d\rho\m d\phi$. \,The restriction to $\,\rho\geq0\,$ will 
not, however, effect the large deviation form of the PDF of $\,\rho$. 
\,In what follows, we shall then simplify the things considering the operator 
$\,\CL^{as}_{\phi\rho}\,$ as acting on functions of $\,\rho\,$ defined on the 
whole line, restricting $\,\rho\,$ to positive values only at the very end.
\vskip 0.3cm

Although $\,\CL^{as}_{\phi\rho}\,$ still does not preserve the subspace of 
functions that depend only $\,\rho$, \,it does preserve the one of functions 
that depend only on $\,\phi$. \,Hence the evolution of $\,\phi\,$ becomes at 
long times independent of that of $\,\rho\,$ although the opposite is not true. 
More specifically, the angle $\,\phi\,$ undergoes at long time a diffusion 
on the circle of circumference $\,\pi\,$ whose generator 
\qq
\CL^{as}_{\phi}\ \,=\,\ 2\alpha\Big[\sin{\phi}
\,\partial_\phi\Big]^2\,+\,2\gamma\,\partial_\phi^2
\label{phi}
\qqq
is obtained by restricting $\,\CL^{as}_{\phi\rho}\,$ to functions
independent of $\,\rho$. Such diffusion process converges exponentially fast 
(we shall find the rate of the exponential convergence below) to a stationary state. 
The stationary density 
\qq
\chi(\phi)\ =\ {\Big[\gamma+\alpha\,\sin^2{\phi}\Big]^{-1/2}
\over\int\limits_0^{\pi}\Big[\gamma+\alpha\,
\sin^2{\varphi}\Big]^{-1/2}d\varphi}\,.
\label{pf}
\qqq
is the unique positive normalized solution of period $\,\pi\,$ of the equation 
\qq
{\CL^{as}_{\phi}}^{\,\dagger}\chi(\phi)\ =\ \Big(2\alpha
\Big[\partial_\phi\,\sin{\phi}\Big]^2\,+\,2\gamma\,\partial_\phi^2\Big)\,
\chi(\phi)\ =\ 0\,.
\nonumber
\qqq

\subsection{Difference of Lyapunov exponents}

\noindent The difference of the Lyapunov exponents may be found now by applying 
the following strategy. Suppose that there exists a smooth periodic 
function $\,f(\phi)\,$ of period $\,\pi\,$ such that
\qq
\CL^{as}_{\phi\rho}\,(\rho+f(\phi))\ =\ \lambda\,=\,{\rm const}.
\label{spt}
\qqq
It follows that
\qq
{d\over dt}\m\Big\langle\rho\Big\rangle\ =\ \lambda\,-\,{d\over dt}\m
\Big\langle f(\phi)\Big\rangle\quad 
\mathop{\longrightarrow}\limits_{t\to\infty}\ \ \lambda
\nonumber
\qqq
with the exponentially fast convergence. We infer then that 
$\,\lambda\,$ is equal to the mean asymptotic rate of change of $\,\rho\,$
that, in turn, is equal to the difference of the Lyapunov exponents. 
Identity (\ref{spt}) may be rewritten in somewhat more explicit form as
\qq
2\alpha\,\sin^2{\phi}\,+\,\beta\,
+\,\gamma\,+\,\CL^{as}_{\phi}\,f(\phi)\ =\ \lambda\,.
\nonumber
\qqq
Integrating the latter equality against $\,\chi(\phi)$, \,we obtain 
the relation
\qq
\lambda\ =\ \int\limits_0^{\pi}\Big[\beta\,+\,\gamma\,+\,
2\alpha\,\sin^2{\varphi}\Big]\,\chi(\varphi)\,d\phi
\label{con}
\qqq
that fixes the value of $\,\lambda$. \,It is easy
to see that Eq.\,\,(\ref{con}) is also a sufficient condition
for the existence of the function $\,f(\phi)\,$ satisfying 
condition (\ref{spt}). Substituting the explicit expression (\ref{pf})
for $\,\chi(\phi)$, \,we infer that
\qq
\lambda_1-\lambda_2\ =\ \beta\,-\,\gamma\,+\,
2{\int\limits_0^{\pi}\Big[\gamma\,+\,\alpha\,\sin^2
{\varphi}\Big]^{1/2}\,d\varphi\over
\int\limits_0^{\pi}\Big[\gamma\,+\,\alpha\,\sin^2
{\varphi}\Big]^{-1/2}\,d\varphi}\,.
\label{dlexp}
\qqq
%or, using also Eq.\,\,(\ref{slexp}), that
%\qq
%\lambda_1\ &=&\ -\,\alpha\,-\,\beta\,-\,\gamma\,+\,
%{\int\limits_0^{\pi}\Big[\gamma\,+\,\alpha\,\sin^2
%{\varphi}\Big]^{1/2}\,d\varphi\over
%\int\limits_0^{\pi}\Big[\gamma\,+\,\alpha\,\sin^2
%{\varphi}\Big]^{-1/2}\,d\varphi}\,,\cr
%\lambda_2\ &=&\ -\,\alpha\,-\,2\beta\,-\,
%{\int\limits_0^{\pi}\Big[\gamma\,+\,\alpha\,\sin^2
%{\varphi}\Big]^{1/2}\,d\varphi\over
%\int\limits_0^{\pi}\Big[\gamma\,+\,\alpha\,\sin^2
%{\varphi}\Big]^{-1/2}\,d\varphi}\,.
%\qqq
Recall that $\,\gamma\geq|\beta|\,$ and $\,2\alpha+2\beta+\gamma\geq|\beta|\,$
and that we have assumed that $\,\alpha\geq 0$. \,It follows that 
$\,\lambda^1-\lambda^2\geq\beta+\gamma\geq0\,$ and at least
one of the last inequalities is sharp unless $\,\beta=\gamma=0$.
Hence $\,\lambda_1>\lambda_2\,$ if $\,\beta+\gamma>0\,$ which is consistent 
with the assumption that, typically, $\,\rho\,$ becomes large for long times. 
The integrals are given by the elliptic functions $\,{\bm K}(k)\,$ and 
$\,{\bm E}(k)\,$ with the modulus $\,k=\sqrt{\alpha\over\alpha+\gamma}\m$:
\qq
{\int\limits_0^{\pi}\Big[\gamma\,+\,\alpha\,\sin^2
{\varphi}\Big]^{1/2}\,d\varphi\over
\int\limits_0^{\pi}\Big[\gamma\,+\,\alpha\,\sin^2
{\varphi}\Big]^{-1/2}\,d\varphi}\ =\ 
(\alpha+\gamma)\,
{{\bm E}(k)\over
{\bm K}(k)}\,.
\nonumber
\qqq

\begin{figure}[ht]
\vskip -0.4cm
  \begin{center}
    \subfigure[\label{fig:p0} $\wp = 0$]{%
      \begin{minipage}{0.26\textwidth}
        \includegraphics[width=\textwidth]{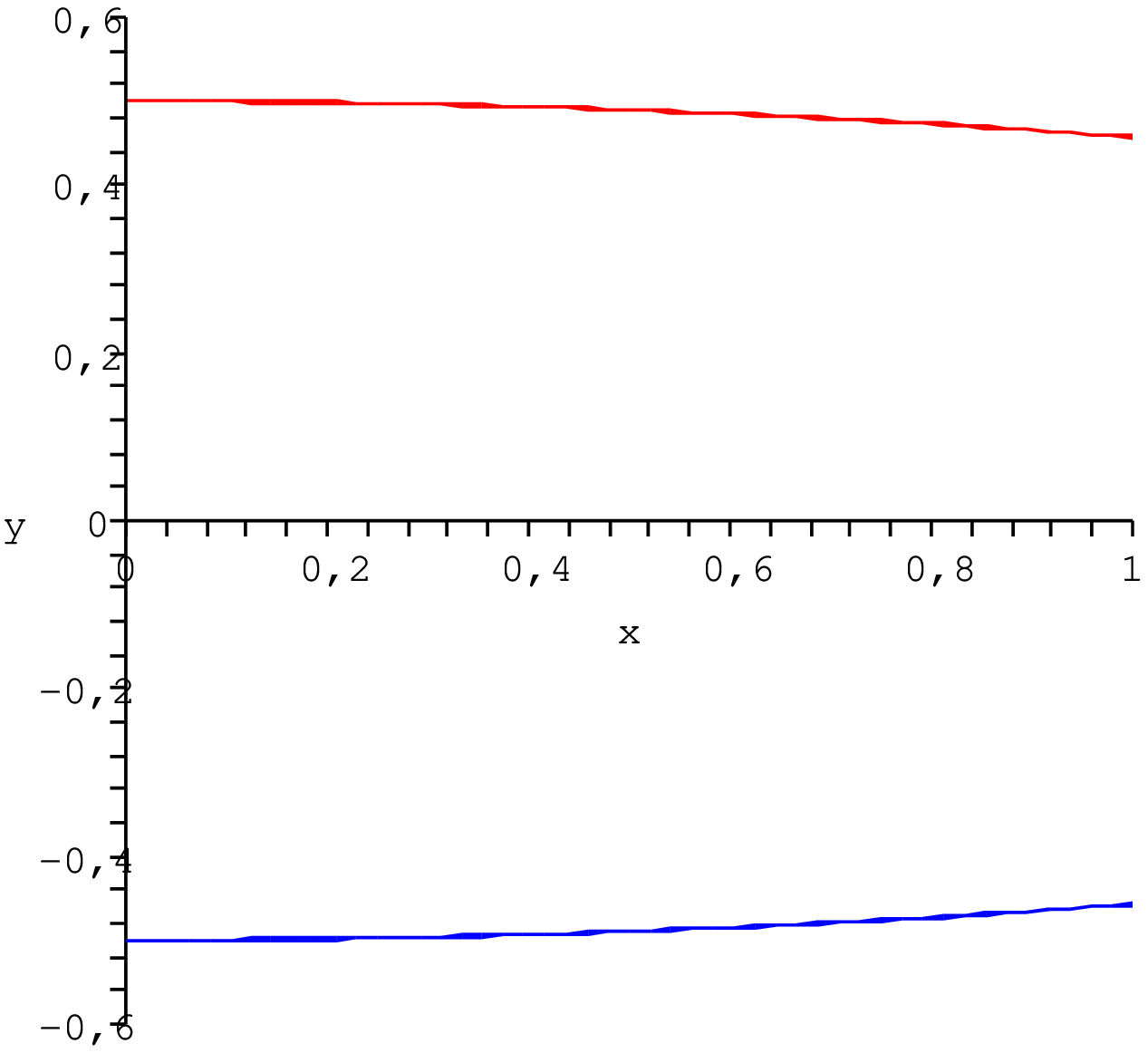}\\
        \vspace{-0.8cm} \strut
        \end{minipage}}
    \hspace{1cm}
    \subfigure[\label{fig:p0.5} $\wp = {1\over2}$]{%
      \begin{minipage}{0.26\textwidth}
        \includegraphics[width=\textwidth]{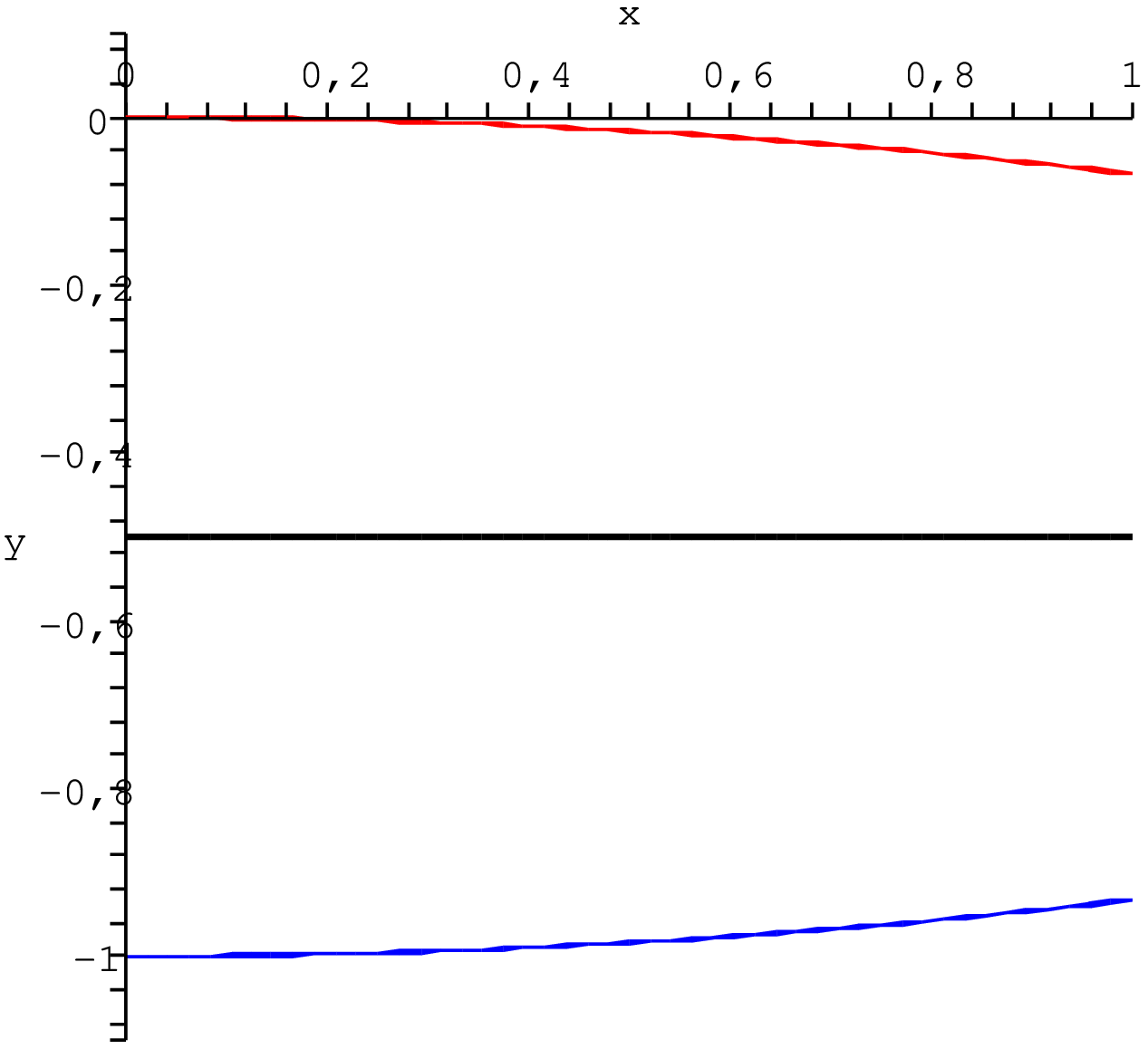}\\
      \vspace{-0.8cm} \strut
        \end{minipage}}
    \hspace{1cm}
    \subfigure[\label{fig:p1} $\wp = 1$]{%
      \begin{minipage}{0.27\textwidth}
      \vspace{0.5cm} 
       \includegraphics[width=\textwidth,height=3.7cm]{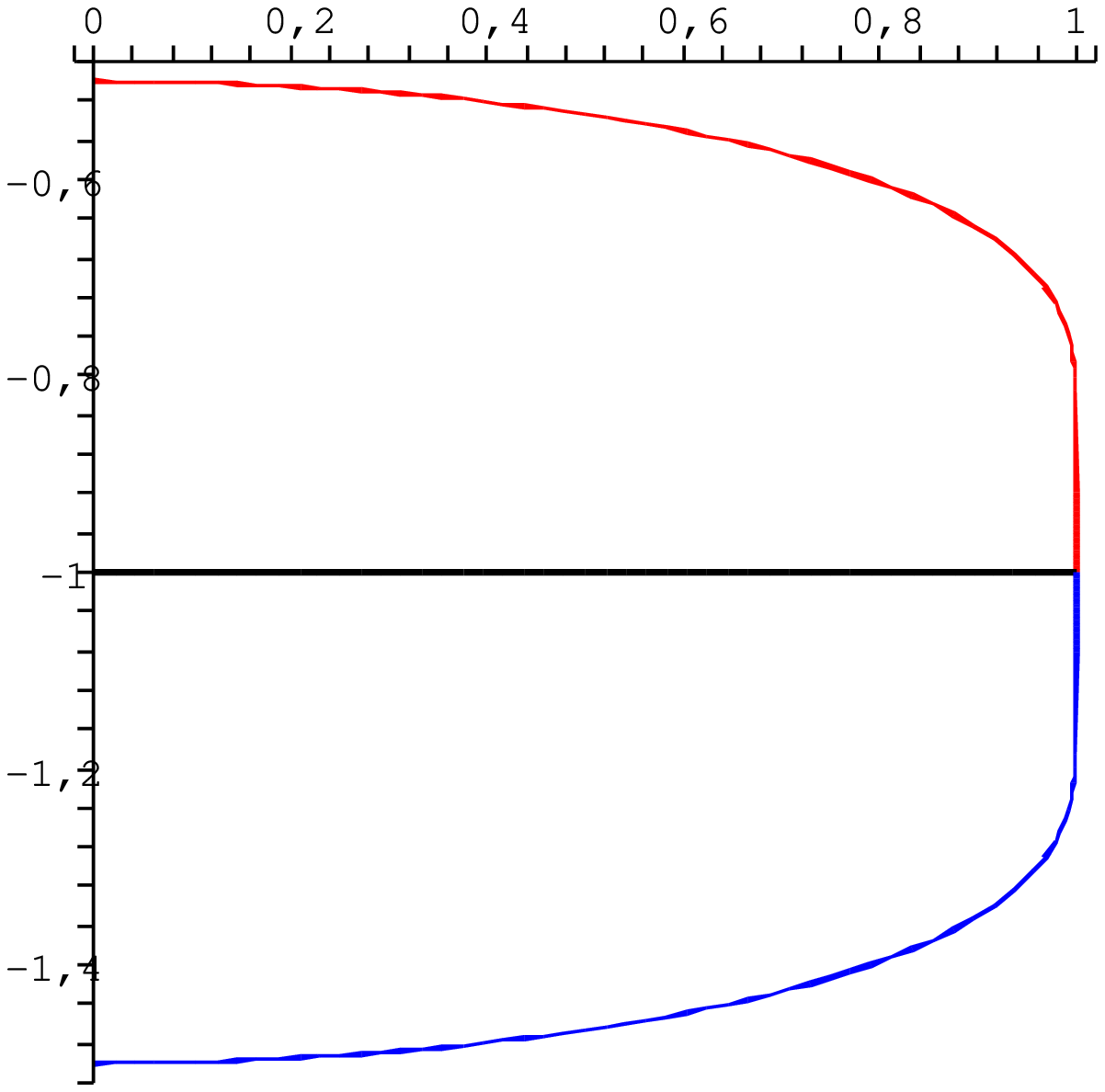}\\
      \vspace{-0.45cm} \strut
        \end{minipage}}
  \end{center}
  \vspace{-0.7cm}
  \caption{\label{fig:fig1} 
Normalized Lyapunov exponents $\,{\lambda_1\tau}\,$ and 
$\,{\lambda_2\tau}\,$ as functions of anisotropy degree
$\,\kappa={|\alpha|\tau}\,$ for three values\\  
\hspace*{1.2cm}of the compressibility degree $\,\wp$.}
\vspace{0.2cm}
\end{figure}

\noindent Fig.\,\ref{fig:fig1} depicts the normalized Lyapunov 
exponents $\,{\lambda_i\tau}\,$ for $\,\tau=(\alpha+\beta+\gamma)^{-1}\,$ as 
functions of the anisotropy degree $\,\kappa={|\alpha|\tau}$, 
\qq
&&{\lambda_1\tau}\, 
=\,-\,1\,+\,{3-2\wp+\kappa\over 2}\,{\bm E(\sqrt{2\kappa\over3-2\wp+\kappa})
\over\bm K(\sqrt{2\kappa\over3-2\wp+\kappa})}\,,\cr
&&{\lambda_2\tau}\,=\, 
1\,-\,2\wp\,-\,{3-2\wp+\kappa\over 2}\,{\bm E(\sqrt{2\kappa\over3-2\wp+\kappa})
\over\bm K(\sqrt{2\kappa\over3-2\wp+\kappa})}\,,
\nonumber
\qqq
for three values of the compressibility degree $\,\wp=0$, $\,\wp={1\over2}\,$ 
and $\,\wp=1$. \,Note that $\,\lambda_1\tau\,$ decreases
and $\,\lambda_2\tau\,$ increases with $\,\kappa\,$ at constant
compressibility degree $\,\wp$, \,with the sum of the two fixed to $\,-2\wp$,
\,see Eq.\,(\ref{sumf}). \,The incompressible system stays always chaotic 
(i.e. with positive top Lyapunov 
exponent) and this is also true for sufficiently small compressibility degree. 
For $\,\wp\,$ slightly below $\,{1\over2}$, \,however, an increase of 
$\,\kappa\,$ may kill chaos. For $\,\wp\geq{1\over 2}\,$ the system is 
never chaotic. For $\,\wp=1$, \,the tendency of anisotropy to bring 
the Lyapunov exponents closer attains its maximum with the two Lyapunov 
exponents coinciding for the extreme anisotropy when $\,S^\omega_t\,$ 
is a diagonal matrix with independent equally distributed entries 
representing independent stretching and contraction along the coordinate 
axes (or when it is the $\,45^o$ rotation of such a matrix).

\subsection{Large deviations for exponents $\,\rho$}

\begin{figure}[ht]
\begin{center}
\vspace{0.2cm}
\mbox{\hspace{0.0cm}\psfig{file=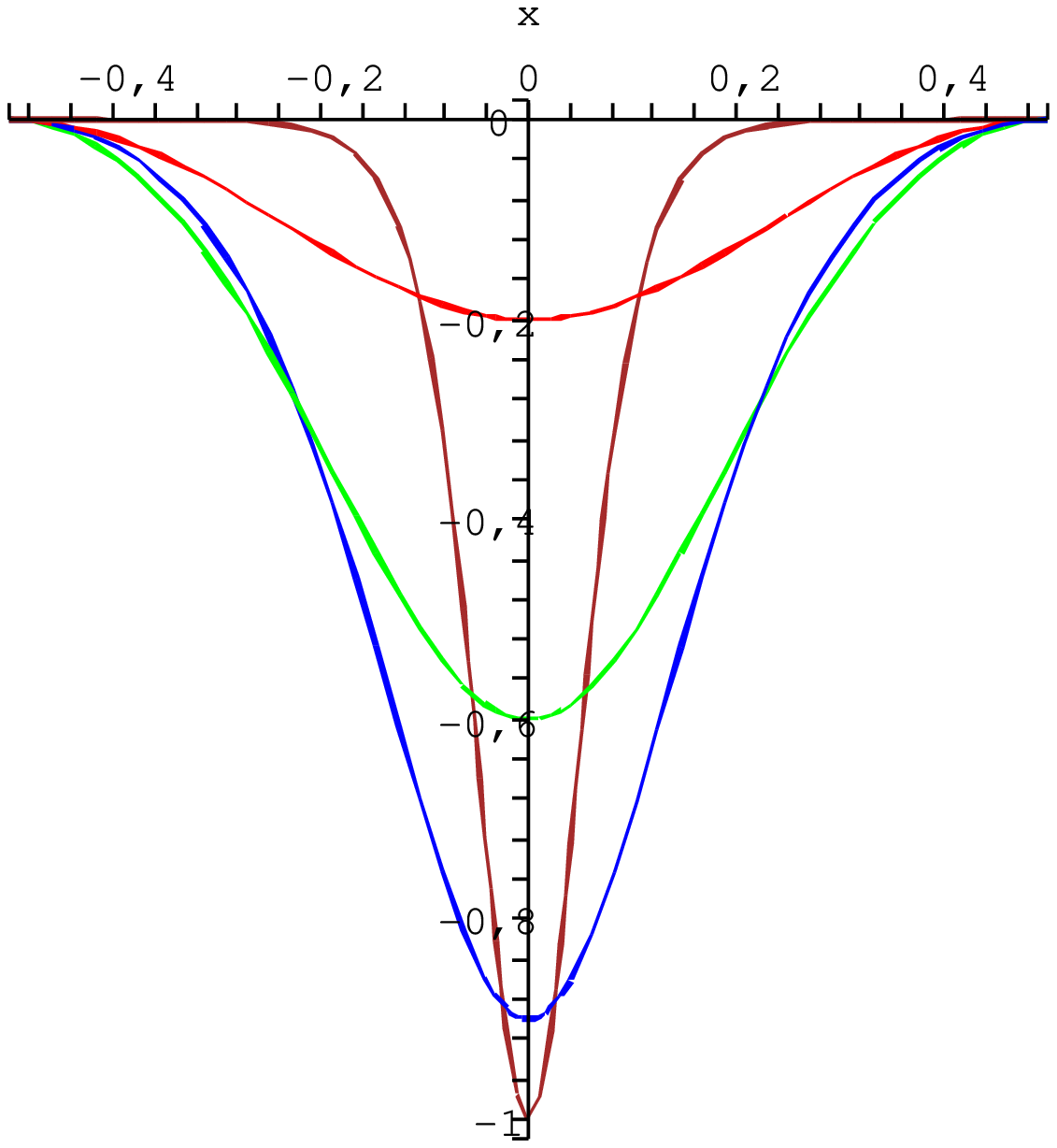,height=6cm,width=8cm}}
\end{center}
\vspace{-0.4cm}
\caption{$V(\phi(u))\,$ 
as a function of $\,{u\over2\bm K(k)}\,$ for, from top to
bottom for $\,u=0$, $\,k^2={\alpha\over\alpha+\gamma}=0.2,\,0.6,\,0.9\
{\rm and}\ 0.99999$.}
\label{fig:fig2}
%\vspace{-0.1cm}
\end{figure}

\noindent The joint PDF of $\,\phi_t\,$ and $\,\rho_t\,$ takes the form 
of the heat kernel
\qq
P_t(\phi,\rho)\ =\ 
\ee^{\,t\,\CL_{\phi\rho}}(0,0;\phi,\rho)\,
\nonumber
\qqq
and for large time $\,t\,$ should be well approximated by the
modified heat kernel
\qq
P^{as}_t(\phi,\rho)\ =
\ \ee^{\,t\,\CL^{as}_{\phi\rho}}(0,0;\phi,\rho)\,.
\nonumber
\qqq
In the latter, the $\,\rho$-contribution may be diagonalized
by the Fourier transform so that we get
\qq
P^{as}_t(\phi,\rho)\ =\  \int\limits_{\CC}
\ee^{-\nu\rho\,+\,t\,(\beta+\gamma)\nu(\nu+1)}
\,\,\ee^{\,t\,\CL_\nu}(0,\phi)\,\,{_{d\nu}\over^{2\pi i}}
\label{ptas}
\qqq
where the integration is over a line $\,Re\,\nu=-1/2\,$ parallel to the imaginary 
axis and
\qq
\CL_\nu\ =\ 2\alpha\Big[\sin{\phi}\,\partial_\phi\,-\,
\nu\,\cos{\phi}\Big]^2\,+\,2\gamma\,\partial_\phi^2
\nonumber
\qqq
is a second order differential operator acting on periodic
functions of period $\,\pi$. \,For ${Re}\,\nu=-1/2$, $\,\CL_\nu\,$ is self-adjoint
with respect to the $\,L^2\,$ scalar product with the measure $\,d\phi$. 
As we shall see below, the contour $\,\CC\,$ of integration in Eq.\,(\ref{ptas})
may be moved to any line parallel to the imaginary axis. Operator $\,\CL_\nu\,$ 
may be viewed as a perturbation of the generator $\,\CL^{as}_\phi\,$ of Eq.\,(\ref{phi}) 
with which it coincides for $\,\nu=0$. \,By a rescaling, a similarity transform and the 
elliptic change of variables
\qq
\phi\ \longmapsto\ u(\phi)\,=\,\int\limits_0^{\phi}
\Big[{\alpha+\gamma\over \gamma\,
+\,\alpha\,\sin^2{\psi}}\Big]^{1/2}d\psi\,,
\nonumber
\qqq
$\CL_\nu\,$ is put into the form of a one-dimensional Schr\"odinger 
operator:
\qq
-\,{_1\over^{2(\alpha+\gamma)}}\,\ee^{-\nu\,h(\phi)}\,\CL_\nu\,
\ee^{\m\nu\,h(\phi)}\ =\ -{d^2\over du^2}\,+\,
\nu(\nu+1)\,V(\phi(u))
\label{Schop}
\qqq
for the function
\qq
h(\phi)\ =\ {_1\over^2}\,\ln\Big[\gamma\,+\,
\alpha\,\sin^2{\phi}\Big]
\nonumber
\qqq
and the attractive potential depicted on \,Fig.\,\ref{fig:fig2}, 
\qq
V(\phi(u))\ =\ {\gamma\over{\alpha+\gamma}}\,-\,{\gamma\over{\gamma
\,+\,\alpha\,\sin^2\phi}}\ =\ -k^2\,{\rm cn}^2(u,k)\ =\ -k^2\,+\,k^2\,{\rm sn}^2(u,k)\,,
\nonumber
\qqq
where $\,{\rm sn}(u,k)\,$ and $\,{\rm cn}(u,k)\,$ are 
the Jacobi elliptic function corresponding to the modulus 
$\,k=\sqrt{{\alpha\over\alpha+\gamma}}$.
The Schr\"odinger operator of Eq.\,(\ref{Schop}) acts on periodic functions 
of $\,u\,$ with the period $\,2\bm K(k)\,$ (that corresponds to the period $\,\pi\,$
in $\,\phi$). \,Up to a constant, it is equal to the Lam\'e integrable operator 
in the Jacobian form \cite{WhittWat,ErdMOT}
\qq
\CH_\nu\ =\ -{d^2\over du^2}\,+\,
\nu(\nu+1)\,k^2\,{\rm sn}^2(u,k)\,.
\nonumber
\qqq
We thus obtain the relation
\qq
P_t^{as}(\phi,\rho)\ =\ \int\limits_{\CC}
\ee^{-\nu\rho\,+\,t\m(2\alpha+\beta+\gamma)\nu(\nu+1)\,+\,\nu(h(0)-h(\phi))}
\,\,\ee^{\,-2(\alpha+\gamma)t\,\CH_\nu}(0,u(\phi))\,\,{_{du(\phi)}\over^{d\phi}}
\,{_{d\nu}\over^{2\pi i}}
\label{ptas1}
\qqq
Note that by the Feynman-Kac formula, for $\,\nu=\nu_1+i\nu_2\,$ with $\,\nu_{1,2}\,$
real, the absolute value of the integrand on the right hand side is bounded by
\qq
\ee^{-\nu_1\rho\,+\,t\m(\beta+\gamma)[\nu_1(\nu_1+1)-\nu_2^2]\,
+\,\nu_1(h(0)-h(\phi)}\,\,\ee^{\,-2(\alpha+\gamma)t\m\left(-{d^2\over du^2}+
[\nu_1(\nu_1+1)-\nu_2^2]V(\phi(u))\right)}(0,u(\phi))
\,\,{_{du(\phi)}\over^{d\phi}}.
\nonumber
\qqq
For $\,\gamma>0$, \,the last expression tends to zero when 
$\,\nu_2\to\pm\infty\,$ (uniformly on bounded intervals of 
$\,\nu_1$) since $\,\beta+\gamma\geq0\,$ and $\,V\,$ is
attractive. \,It follows then that the contour of integration $\CC\,$ on 
the right hand side of Eq.\,(\ref{ptas1}) may be, as announced, moved to 
any line parallel to the imaginary axis. Let us consider the spectral 
decomposition
\qq
\ee^{-2(\alpha+\gamma)t\,\CH_\nu}\ =\ \sum\limits_{n=0}^\infty\ee^{-2(\alpha+\gamma)t\,
E_{\nu,n}}
\,|\Psi_{\nu,n}\rangle\langle\Psi_{\nu,n}|
\label{spdec}
\qqq
with the eigenvalues $\,E_{\nu,n}=E_{\nu,n}(k^2)\,$ of $\,\CH_\nu\,$ ordered in a 
non-decreasing way for $\,\nu\,$ real. 
%The spectral decomposition (\ref{spdec}) implies the following
%expansion for the heat kernel of $\,\CL_\nu$:
%\qq
%\ee^{\,t\,\CL_\nu}(0;\phi)\ =\ \ee^{\m\nu\,h(0)}
%\,\sum\limits_{n=0}^\infty\ee^{\,-2(\alpha+\gamma)t\,[-\nu(\nu+1)k^2+E_{\nu,n}(k^2)]}\,\,
%\langle u(0)|\Psi_{\nu,n}\rangle\langle
%\Psi_{\nu,n}|u(\phi)\rangle\,\,\ee^{-\nu\,h(\phi)}
%\,\,{_{du(\phi)}\over^{d\phi}}\,.
%\label{spex}
%\qqq
For large $\,t$, \,the dominant contribution comes from the ground state 
$\,\Psi_{\nu,0}\,$ of $\,\CH_\nu$. \,In particular, for the vanishing 
coupling constant, $\,\Psi_{0,0}=(2\bm K(k))^{-1/2}$, $\,E_{0,0}=0$, 
$\,E_{0,1}=E_{0,2}=\pi^2\bm K(k)^{-2}\,$ and 
$\,\ee^{\,t\,\CH_0}(0;u(\phi))\,$ converges at long times to 
$\,(2\bm K(k))^{-1}\,$ with the exponential rate equal to 
$\,2\pi^2(\alpha+\gamma)
\bm K(k)^{-2}$. \,Note that this rate goes to zero when $\,\gamma\,$
tends to zero since the half-period $\,\bm K(k)\,$ diverges in this limit.
\vskip 0.3cm

Insertion of the expansion (\ref{spdec}) into the expression 
(\ref{ptas1}) permits to extract the large deviation form of the PDF 
of $\,\rho_t\,$ from the ground state contribution:
\qq
\int\limits_0^{\pi} P_t^{as}(\phi,\rho)\,d\phi
\ \ \propto\ \ \int\limits_{\CC}
\ee^{-t\m\left[\nu\rho/t\,-\,(2\alpha+\beta+\gamma)\nu(\nu+1)
\,+\,2(\alpha+\gamma)\,E_{\nu,0}(k^2)\right]}
\,\,{_{d\nu}\over^{2\pi i}}
\ \ \propto\ \ \ee^{-t\m H_\rho(\rho/t)}\ \ 
\label{approxt}
\qqq
with the rate function
\qq
H_\rho(\rho/t)\ =\ \max\limits_\nu
\Big[\nu\rho/t\,-\,(2\alpha+\beta+\gamma)\nu(\nu+1)\,+
\,2(\alpha+\gamma)\,E_{\nu,0}(k^2)\Big]
\label{rhold}
\qqq
defined for $\,\rho/t>0$. \,We shall denote by $\,\nu_{max}\,$ the value 
of $\,\nu\,$ where the maximum is attained
(the maximum over real $\,\nu\,$ corresponds to a minimum of the real part
along the axis parallel to the imaginary one). 
Note that only the ground 
state energy of operator $\,\CH_\nu\,$ contributes to the rate function 
$\,H_\rho$. \,The ground state wave function enters the prefactors in 
the long time asymptotics of the PDF of $\,\rho$. Let us check that when 
$\,t\to\infty\,$ then $\,\rho/t\,$ concentrates at the value equal to the 
difference of the Lyapunov exponents as given by Eq.\,\,(\ref{dlexp}). 
To this end, we must find the minimum of $\,H_\rho(\rho/t)$. 
\,The stationarity condition implies the equations
\qq
\nu=0\,,\qquad \rho/t\,=\,(2\alpha+\beta+\gamma)(2\nu+1)\,
-\,2(\alpha+\gamma)\,\partial_\nu E_{\nu,0}(k^2)
\label{statio}
\qqq
and the minimizing value of $\,\rho/t\,$ is given by the relation
\qq
(\rho/t)_{min}\,=\,2\alpha\,+\,\beta\,+\,\gamma\,
-\,2(\alpha+\gamma)\,\partial_\nu E_{\nu,0}(k^2)|_{\nu=0}\,.
\nonumber
\qqq
The derivative of the ground state energy may be calculated
by the first order perturbation theory:
\qq
&\displaystyle{\partial_\nu E_{\nu,0}|_{\nu=0}\ =\ \langle\Psi_{0,0}
|k^2\m{\rm sn}^2(\m\cdot\m,k)|\Psi_{0,0}\rangle\ =\ k^2\,
+\,{1\over 2\bm K(k)}\int\limits_0^{2\bm K(k)}V(\phi(u)\,du}&\cr 
&\displaystyle{=\ 1\,-\,{1-k^2\over2\bm K(k)}\int\limits_0^\pi
\Big[1-k^2\,\cos^2\phi]^{-3/2}\,d\phi\ =\ 1\,-\,{\bm E(k)\over\bm K(k)}}&
\nonumber
\qqq 
where the last equality follows from the elliptic identity 2.584.37 in 
\cite{GrR}. We obtain this way the relation
\qq
(\rho/t)_{min}\ =\ \beta\,-\,\gamma\,+\,
2(\alpha+\gamma)\,{\bm E(k)\over\bm K(k)}
\nonumber
\qqq
which agrees with  Eq.\,(\ref{dlexp}) for $\,\lambda_1-\lambda_2\equiv\lambda$.
\vskip 0.2cm

A closed expression for the second derivative of the rate function 
$\,H_\rho\,$ may be obtained by differentiating
twice the defining relation (\ref{rhold}):
\qq
H''_\rho(\rho/t)\ =\ {d\nu_{max}\over d(\rho/t)}
\ =\ {1\over 2\left(2\alpha+\beta+\gamma\,-\,(\alpha+\gamma)\,
\partial_\nu^2E_{\nu,0}(k^2)|_{\nu=\nu_{max}}\right)}
\label{h''}
\qqq
since $\,\nu_{\max}\,$ is related to $\,\rho/t\,$ by the second of 
Eqs.\,(\ref{statio}) whose differentiation leads to the last equality.
In particular,
\qq
H''_\rho(\lambda)\ =\ {d\nu_{max}\over d(\rho/t)}\Big|_{\hspace{-0.05cm}{
\lambda}}\ =\ {1\over 2\left(2\alpha+\beta+\gamma\,-\,(\alpha+\gamma)\,
\partial_\nu^2E_{\nu,0}(k^2)|_{\nu=0}\right)}
\label{h''ll}
\qqq
By the second order perturbation expansion,
\qq
\partial_\nu^2E_{\nu,0}|_{\nu=0}\,&=&\,2\partial_\nu E_{\nu,0}|_{\nu=0}\ -\ 
2\sum\limits_{n=1}^\infty E_{0,n}^{-1}\,|\langle \Psi_{0,0}
|k^2\m{\rm sn}^2(\m\cdot\m,k)|\Psi_{0,n}\rangle|^2\cr 
&=&\,2-2{\bm E(k)\over\bm K(k)}\,-\,{k^4\bm K(k)^2\over\pi^2}
\sum\limits_{n=1}^\infty n^{-2}\m f_n(k)^2
\label{2ndder}
\qqq
with the Fourier coefficients $\,f_n(k)=\int\limits_{-1}^1\cos(\pi nx)\,
{\rm sn}^2(\bm K(k)\m x,k)\,dx\m$. $\,H''_\rho(\lambda)\,$ 
is equal to the inverse variance of the normal random variable obtained 
by the central limit $\,\,\lim\limits_{t\to\infty}\,(\rho_t-\lambda)
/\sqrt{t}$. 
\vskip 0.2cm

Combining Eqs.\,(\ref{rhold}) and (\ref{rld}), one obtains the following 
form of the large deviations rate functions for the stretching exponents:
\qq
H({\rho_1/t},{\rho_2/t})\,&=&\,{({\rho_1\over t}+{\rho_2\over t}
+2\alpha+3\beta+\gamma)^2\over 4(2\alpha+3\beta+\gamma)}\cr\cr
&&+\,\max\limits_\nu\Big[\nu({_{\rho_1}\over^t}-{_{\rho_2}\over^t})
-(2\alpha\beta+\gamma)\nu(\nu+1)+2(\alpha+\gamma)\,
E_{\nu,0}({_\alpha\over^{\alpha+\gamma}})\Big].\, 
\label{LDRF}
\qqq
for $\,\rho_1>\rho_2$. \,When $\,\alpha=0\,$ (the isotropic case) then 
$\,E_{\nu,0}=0\,$ and the large deviation rate function (\ref{LDRF}) 
reduces to the one given by the two-dimensional version of expression 
(\ref{LDH}).

\subsection{Large deviations for exponents $\,\bm\eta$}

\noindent Similarly as in the isotropic case, the PDF of the stretching 
exponents $\,{\bm\eta}_t\,$ along the unstable flag may be obtained 
from Eq.\,(\ref{tWf}) by considering the Iwasawa decomposition
\qq
\tilde W\ =\ \left(\matrix{\cos{\phi\over 2}&\sin{\phi\over 2}\cr
-\sin{\phi\over 2}&\cos{\phi\over 2}}\right)
\left(\matrix{\ee^{\eta_1}&0\cr0&\ee^{\eta_2}}\right)
\left(\matrix{1&x\cr 0&1}\right)\,.
\label{IW}
\qqq
of the solutions of the linear stochastic equation (\ref{ldes})
with the initial condition given by a (random) rotation matrix.
In the parametrization (\ref{IW}) and upon the substitution 
$\,{1\over2}(\eta_1+\eta_2)=r$, $\,\eta_1-\eta_2=\eta$, the generator 
$\,\CL\,$ of the tangent process takes the form:
\qq
\CL\ &=&\ {_1\over^2}(2\alpha+\beta+\gamma)
\Big[{_1\over^2}\partial_r^2\,+\,2\cos^2\phi\,\partial_\eta^2
+\,2\sin^2\phi\,\partial_\phi^2\,+\,2\sin^2\phi\,\ee^{-2\eta}\partial_x^2
\,-\,2\sin(2\phi)\,\partial_\eta\partial_\phi
\cr
&&+2\sin(2\phi)\,\ee^{-\eta}\partial_\eta\partial_x\,
-\,4\sin^2\phi\,\ee^{\-\eta}\partial_\phi\partial_x
\,+\,2\sin^2\phi\,\partial_\eta\,
+\,\sin(2\phi)\,\partial_\phi\,-\,2\sin(2\phi)
\,\ee^{-\eta}\partial_x\Big]
\cr
&&{_1\over^2}(\beta+\gamma)\Big[2\sin^2\phi\,\partial_\eta^2\,
-\,2\sin^2\phi\,\partial_\phi^2\,+\,2\cos^2\phi\,
\ee^{-2\eta}\partial_x^2\,+\,2\sin(2\phi)\,\partial_\eta\partial_\phi
\,-\,2\sin(2\phi)\,\ee^{-\eta}
\partial_\eta\partial_x
\cr
&&-\,4\cos^2\phi\,
\ee^{-\eta}\partial_\phi\partial_x\,+\,2\cos^2\phi\,
\partial_\eta\,-\,\sin(2\phi)\,\partial_\phi
\,+\,2\sin(2\phi)\,\ee^{-\eta}\partial_x\Big]\,+\,2\gamma\,
\partial_\phi^2\cr
&&+\,{_1\over^2}\beta\m\partial_r^2\,
-\,{_1\over^2}(2\alpha+3\beta+\gamma)
\m\partial_r\,.
\nonumber
\qqq
It is self-adjoint in the $\,L^2\,$ scalar product with the measure 
$\,\ee^{-2r+\eta}d\phi\m dr\m d\eta\m dx$. 
\,In the action on functions independent on $\,x$, \,$\,\CL\,$ reduces 
to the sum $\,\CL_r+\CL_{\phi\eta}\,$ with 
$\,\CL_r\,$ given by Eq.\,(\ref{clr}) and
\qq
\CL_{\phi\eta} \ =\ 2\alpha\Big[\sin{\phi}
\,\partial_\phi\,-\,\cos{\phi}\,
\partial_\eta\Big]^2\,+\,2\gamma\,\partial_\phi^2
\,+\,(\beta+\gamma)\Big[\partial_\eta^2\,+\,
\partial_\eta\Big].
\label{phieta}
\qqq
Note that the operator $\,\CL_{\phi\eta}\,$ has the form identical 
to the asymptotic form $\,\CL^{as}_{\phi\rho}\,$ of $\,\CL_{\phi\rho}$, 
\,see Eq.\,(\ref{gg}). In principle, it acts now on functions 
of $\,\phi\,$ with period $\,4\pi\,$ but it preserves the subspace 
of functions with period $\,\pi$. \,The evolution of $\,r\,$ decouples 
from that of $\,\phi\,$ and $\,\eta\,$ leading to the PDF (\ref{rld}). 

\begin{figure}[ht]
\begin{center}
\vspace{0.4cm}
\mbox{\hspace{0.0cm}\psfig{file=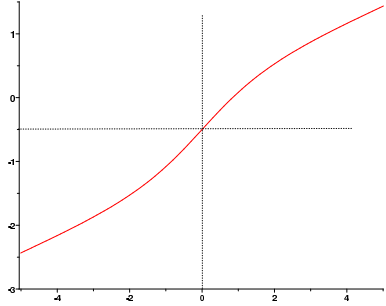,height=4cm,width=5cm}}
\end{center}
\vspace{-0.1cm}
\caption{$\nu_{max}\,$ as a function of $\,(\eta/t)\tau\,$
for $\,\kappa=0.8\,$ and $\,\wp=0.9$.}
\label{fig:fig3}
\vspace{0.1cm}
\end{figure} 
\vskip 0.4cm

As for the joint time $\,t\,$ PDF of $\,\phi\,$ and $\,\eta$, \,it is 
related to the heat kernel of $\,\CL_{\phi\eta}\,$ by the equality
\qq
P_t(\phi,\eta)\ =\ \int\limits_0^\pi \ee^{\m t\m\CL_{\phi\eta}}
(\phi_0,0;\phi,\eta)\m\,\chi(\phi_0)\,d\phi_0\,
\label{ptpe}
\qqq
obtained from Eq.\,(\ref{tWf}). Indeed, it follows from the relation
relation (\ref{mdyc}) that the probability measure $\,\chi(dF)\,$ on
the Grassmannian $\,\Gr\,$ of the 1-dimensional subspaces spanned
by vectors $\,(\cos{\phi_0\over2},-\sin{\phi_0\over2})\,$ 
has to be proportional to $\,\chi(\phi_0)\,d\phi_0\,$ where the function 
$\,\chi(\phi)\,$ is the stationary density (\ref{pf}) of the angle $\,\phi$. 
Since $\,\chi(\phi)\,$ is periodic with period $\,\pi\,$ (as a consequence 
of the invariance of the law of the tangent process with respect to the 
rotations by $\,90^o$), so must be $\,P_t(\,\cdot\,,\eta)\,$ and we could
restrict the $\,\phi_0$-integration in Eq.\,(\ref{ptpe}) to the
interval $\,[0,\pi[$. \,The rest of the analysis of the large 
deviations of $\,\eta\,$ runs as for the large deviations of $\,\rho$. 
\,As the result, the large deviation rate function for $\,\bm\eta\,$ has 
the same functional form (\ref{LDRF}) as that for $\,\bm\rho\,$ except 
of the absence of the restriction $\,\eta_1>\eta_2$. \,In particular,
the rate function $\,H_\eta\,$ of the difference $\,\eta=\eta_1-\eta_2\,$
of the stretching exponents coincides with the function $\,H_\rho\,$
as defined by Eq.\,(\ref{rhold}) on the whole real line rather than
on the half line:
\qq
H_\eta(\eta/t)\ =\ \max\limits_\nu
\Big[\nu\eta/t\,-\,(2\alpha+\beta+\gamma)\nu(\nu+1)\,+
\,2(\alpha+\gamma)\,E_{\nu,0}(k^2)\Big].
\label{etald}
\qqq
Note that $\,H_\eta(\eta/t)+\eta/(2t)\,$ is then an even function of 
$\,\eta/t$. \,Indeed, the values of $\,\nu_{max}\,$ at which the maximum 
on the right hand side of Eq.\,(\ref{etald}) is attained for $\,\eta/t\,$
and for $\,-\eta/t\,$ are related by the reflection around $\,\nu=-1/2$. 
\,They are smaller than $\,-1\,$ for $\,\eta/t<-\lambda$, 
lie in between $\,-1\,$ and $\,0\,$ for $\,-\lambda<\eta/t<\lambda\,$ 
and are positive for $\,\eta/t>\lambda$. \,Fig.\,\ref{fig:fig3}
presents the graph of the maximizing $\,\nu_{max}\,$ as a function
of $\,(\eta/t)\tau\,$ for $\,\kappa=0.8\,$ and $\,\wp=0.9$.
\,We obtain this way the fluctuation relation that compares
the rate function for opposite values of $\,\eta/t\m$:
\qq
H_\eta(-\eta/t)\ =\ H_\eta(\eta/t)\,+\,\eta/t\,.
\nonumber
\qqq
It resembles the Evans-Cohen-Morriss-Gallavotti-Cohen
relation (\ref{GC}) but is different from it. It states, in particular,
that although $\,\eta/t\,$ may take negative values, the probability of
such events is exponentially suppressed for large $\,t$.

\subsection{Properties of the rate function $\,H_\eta$}

\begin{figure}[ht]
\begin{center}
%\vspace{-0.1cm}
\mbox{\hspace{0.0cm}\psfig{file=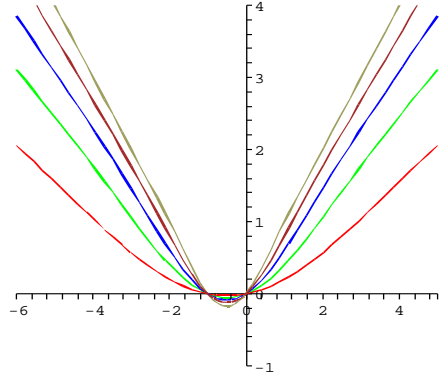,height=6cm,width=7cm}}
\end{center}
\vspace{-0.5cm}
\caption{$E_{\nu,0}(k^2)\,$ as a function of $\,\nu\,$ from bottom to top
for $\,k^2=0.2,\,0.4,\,0.6,\,0.8\,$ and $\,1$.}
\label{fig:fig4}
\vspace{0.3cm}
\end{figure}

\noindent The large deviations rate function $\,H_\eta\,$ is related to
the ground state energy $\,E_{\nu,0}(k^2)\,$ of the Lam\'e operator 
$\,\CH_\nu\,$ for $\,k^2\equiv{\alpha\over\alpha+\gamma}\,$ by the formula 
(\ref{etald}). A lot is known about the eigenvalues and eigenfunctions of 
$\,\CH_\nu$.  \,The power series expansions for the eigenfunctions in an 
appropriate parametrization may be obtained by solving a recursion relation 
\cite{WhittWat} or by diagonalizing tridiagonal matrices \cite{ErdMOT}. 
For $\,\nu\,$ a positive integer and the lowest $\,2\nu+1\,$ eigenvalues, 
the corresponding matrices become finite and one obtains as the eigenfunctions 
the ``Lam\'e polynomials''. The Lam\'e operator may be diagonalized
(for general quasi-momenta) by the Bethe Ansatz \cite{EtingKir,GawFal} that in 
this case goes back to the work of Hermite \cite{WhittWat} in the second half
of the nineteenth century. A simple Maple program computes 
$\,E_{\nu,0}(k^2)\,$ \cite{Volk}. \,We also used a C program 
to compute $\,E_{\nu,0}\,$ by solving directly the eigenvalue equation. 
Fig.\,\ref{fig:fig4} presents the graph of $\,E_{\nu,0}(k^2)\,$ 
as a function of $\,\nu\,$ for $\,k^2=0.2,\,0.4,\,0.6,\,0.8\,$ and
$\,1$.

\begin{figure}[ht]
\begin{center}
\vspace{0.1cm}
\mbox{\hspace{0.0cm}\psfig{file=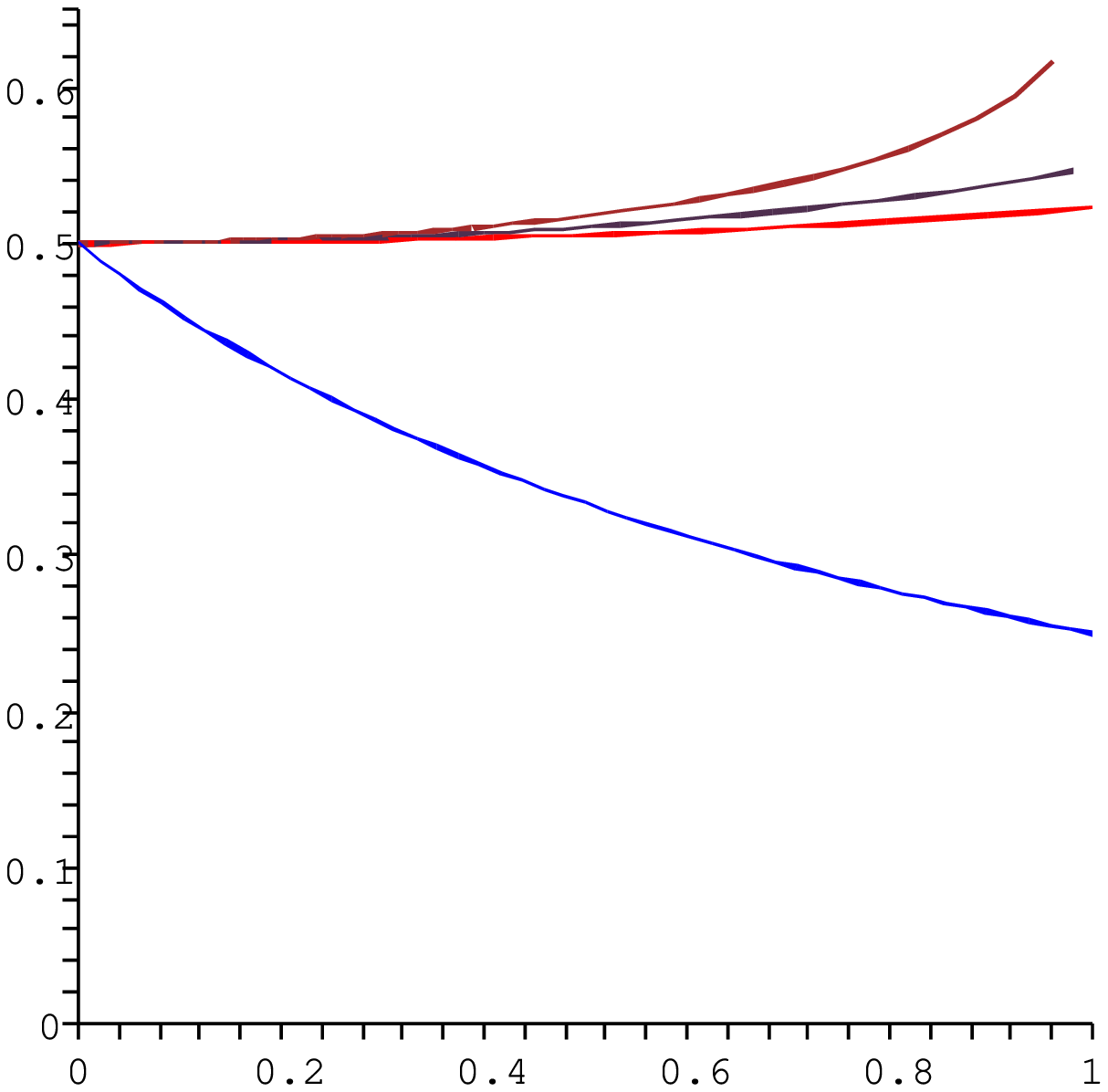,height=5.5cm,width=7cm}}
\end{center}
\vspace{-0.2cm}
\caption{From bottom to top: $\,{1\over\tau}\m H''_\eta(\pm\infty)\,$
(the lower curve) and
${1\over\tau}\m H''_\eta(\lambda)\,$ for $\,\wp=0,\,0.7\,$ and $\,1\,$
as functions of the \\ \hspace*{1.2cm}anisotropy degree $\,\kappa$.}
\label{fig:fig5}
%\vspace{-0.1cm}
\end{figure}

\noindent For large values of $\,\nu\,$ \cite{Volk}, 
\qq
E_{\nu,0}(k^2)\,=\,|k\nu|-{_1\over^4}(1+k^2)\ +\ \CO(|\nu|^{-1})\,.    
\nonumber
\qqq
This leads to the following large $\,|\rho|/t\,$ behavior of $\,H_\eta\m$:
\qq
H_\eta(\eta/t)\ =\ -{\eta\over{2t}}\,+\,{\left({|\eta|\over2t}
+\sqrt{\alpha(\alpha+\gamma)}\right)^2\over2\alpha+\beta+\gamma}\,
-\,{2\alpha+\gamma\over4}\,+\,\CO(|\eta|^{-1})\,.
\nonumber
\qqq
Note that the left and right asymptotes are parabolas 
(displaced with 
each other along the horizontal axis) and that
\qq
H''_\eta(\pm\infty)\ =\ {1\over 2(2\alpha+\beta+\gamma)}\,,
\nonumber
\qqq
in agreement with Eq.\,(\ref{h''}) since $\,\partial_\nu^2E_{\nu,0}\,$
tends to zero at large $\,|\nu|$. \,Recall that we have calculated 
the central-limit inverse variance $\,H''_\eta(\lambda)\,$ before, 
see Eq.\,(\ref{h''ll}), so that around the minimum, 
\qq
H_\eta(\eta/t)\ =\ {1\over4}\,{({\eta\over t}-\lambda)^2\over
2\alpha+\beta+\gamma\,-\,(\alpha+\gamma)
\partial_\nu^2E_{\nu,0}(k^2)|_{\nu=0}}\,\,+\,
\CO(|{_\eta\over^t}-\lambda|^3)
\nonumber
\qqq
with $\,\partial_\nu^2E_{\nu,0}\,$ given by Eq.\,(\ref{2ndder}).
The difference between $\,H''_\eta(\pm\infty)\,$ 
and $\,H''_\eta(\lambda)\,$ attests to the non-Gaussian character of 
the large deviations of $\,\eta$. \,Fig.\,\ref{fig:fig5} presents 
the behavior of $\,{1\over\tau}\m H''_\eta(\pm\infty) 
\,$ and of $\,{1\over\tau}\m H''_\eta(\lambda)\,$ for $\,\wp=0,\,0.7$ and $1$
\,as  functions of the anisotropy degree $\,\kappa={\alpha\tau}$.

\vskip 0.2cm

\begin{figure}[ht]
  \begin{center}
    \subfigure{%
      \begin{minipage}{0.34\textwidth}
        \includegraphics[width=\textwidth]{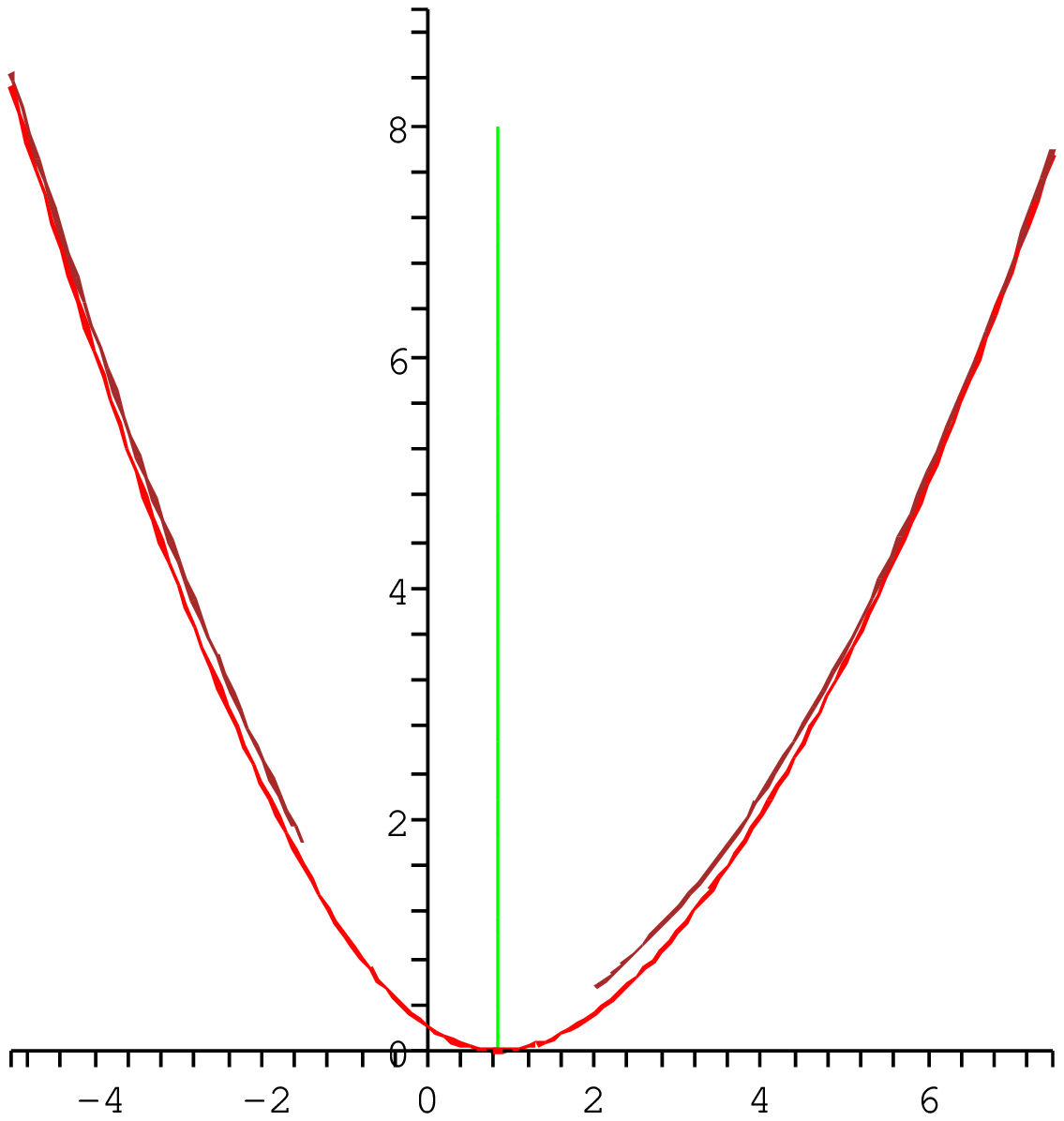}\\
        \vspace{-0.8cm} \strut
        \end{minipage}}
    \hspace{2cm}
    \subfigure{
      \begin{minipage}{0.34\textwidth}
        \includegraphics[width=\textwidth]{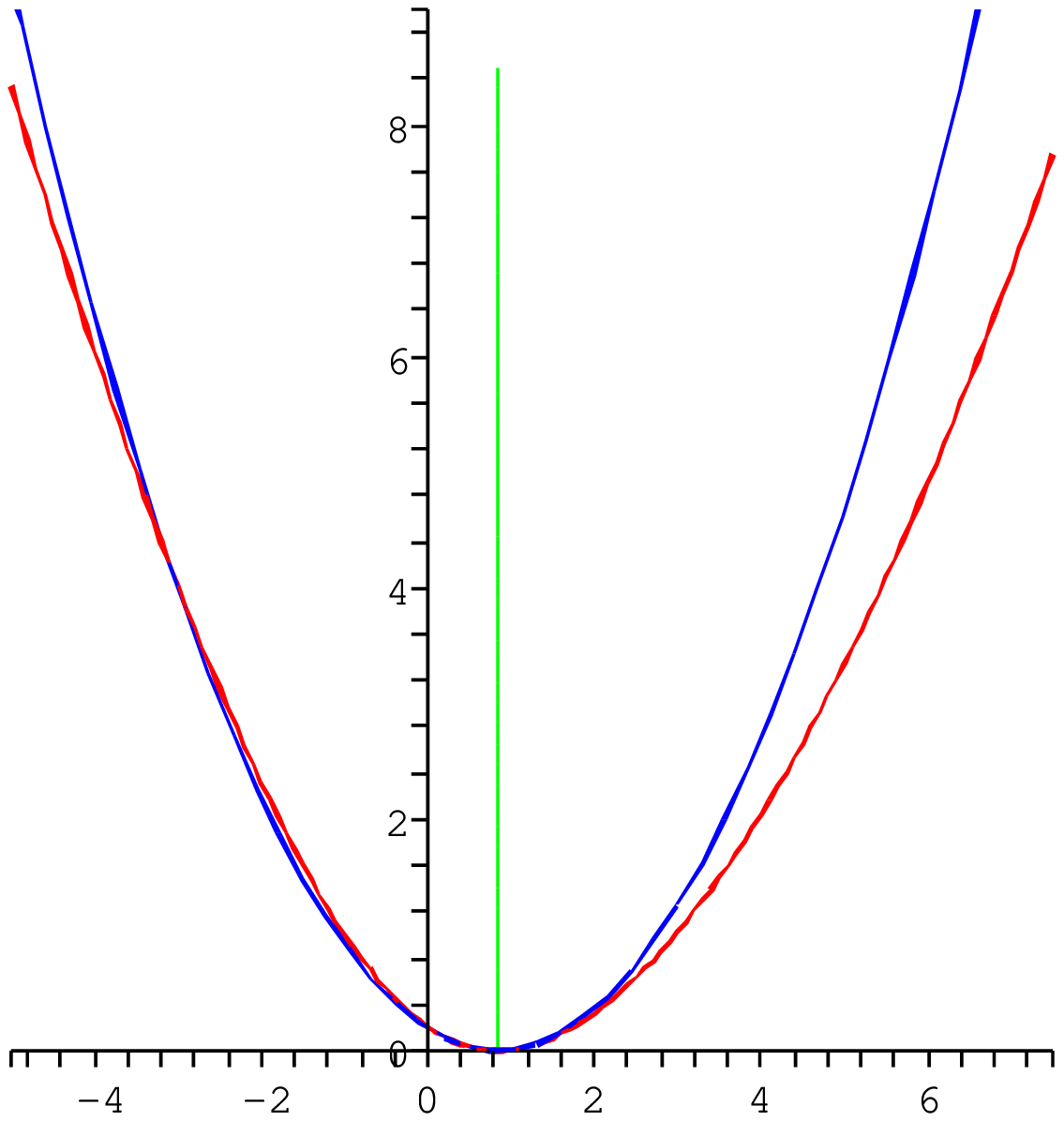}\\
      \vspace{-0.8cm} \strut
        \end{minipage}}
      \end{center}
  \vspace{-0.2cm}
  \caption{\label{fig:fig6} 
$\tau\m H_\eta(x/\tau)\,$ as a function of $\,x\,$ with the large
$\,\eta\,$ asymptotes on the left figure and the quadratic \\ 
\hspace*{1.07cm} approximation around the minimum (the curve tighter 
for large values) for $\,\kappa={0.8}\,$ and $\,\wp={0.9}$.\\ 
\hspace*{1.07cm} The vertical lines correspond to  
$\,\eta/t=\lambda$.}
  \vspace{0.1cm}
\end{figure}

\noindent The first quantity diminishes with growing $\,\kappa\,$ from the 
isotropic value $\,{1\over2}\,$ for $\,\kappa=0\,$ to the extremely 
anisotropic one equal to $\,{1\over 4}\,$ for $\,\kappa=1\,$ whereas 
the second one increases starting from the same initial value. 
It is plausible that for $\,\wp=1$, $\,{1\over\tau}\m H''_\eta(\lambda)\,$ 
diverges in the limit $\,\kappa\to1$. \,We infer that
the undimensionalized central-limit covariance $\,{1\over\tau}\m 
H''_\eta(\lambda)\,$ increases with anisotropy and that the probability 
of large values of $\,\eta/t\,$ decreases slower than if the rate function 
$\,H_\eta\,$ were quadratic with the $\,H''_\eta\,$ equal to its 
central-limit value. \,Fig.\,\ref{fig:fig6} represents the graph 
$\,H_\eta\,$ together with the large value asymptotes and the quadratic 
approximation near the minimum for $\,\kappa={0.8}\,$ and $\,\wp={0.9}$.  

\subsection{Time scales of the large deviations regime}

\noindent It is interesting to look at the time scales at which the large
deviation regime for the difference $\,\rho\,$ or $\,\eta\,$ of 
the stretching exponents sets in. There were three approximations involved 
in reducing the exact PDFs to their large deviation form.
Let us analyze them one by one.
\vskip 0.3cm

The first approximation consisted in replacing the PDF $\,P_t(\phi,\rho)\,$ 
by $\,P^{as}_t(\phi,\rho)\,$ involving the asymptotic form 
$\,\CL^{as}_{\phi\rho}\,$ of the generator $\,\CL_{\phi\rho}$.
The asymptotic form should set in exponentially fast in time with the 
rate given by the difference $\,\lambda\,$ of the Lyapunov exponents. 
This approximation becomes exact when one analyzes the difference 
$\,\eta\,$ of the stretching exponents along the unstable flags.  

\begin{figure}[t]
  \begin{center}
    \subfigure[\label{fig:ga}]{%
      \begin{minipage}{0.3\textwidth}
        \includegraphics[width=\textwidth]{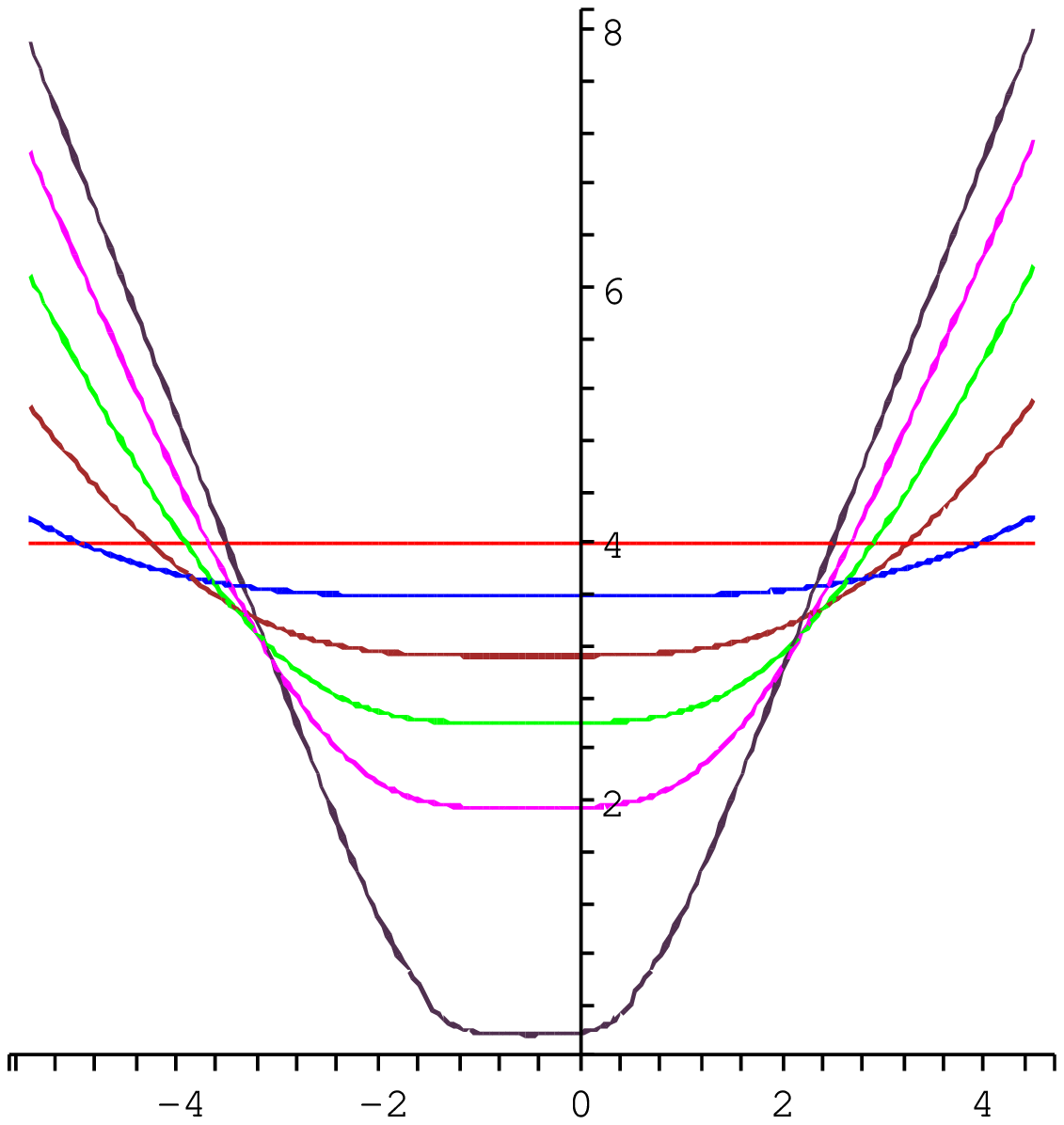}\\
        \vspace{-0.8cm} \strut
        \end{minipage}}
    \hspace{2cm}
    \subfigure[\label{fig:gb}]{%
      \begin{minipage}{0.3\textwidth}
        \includegraphics[width=\textwidth]{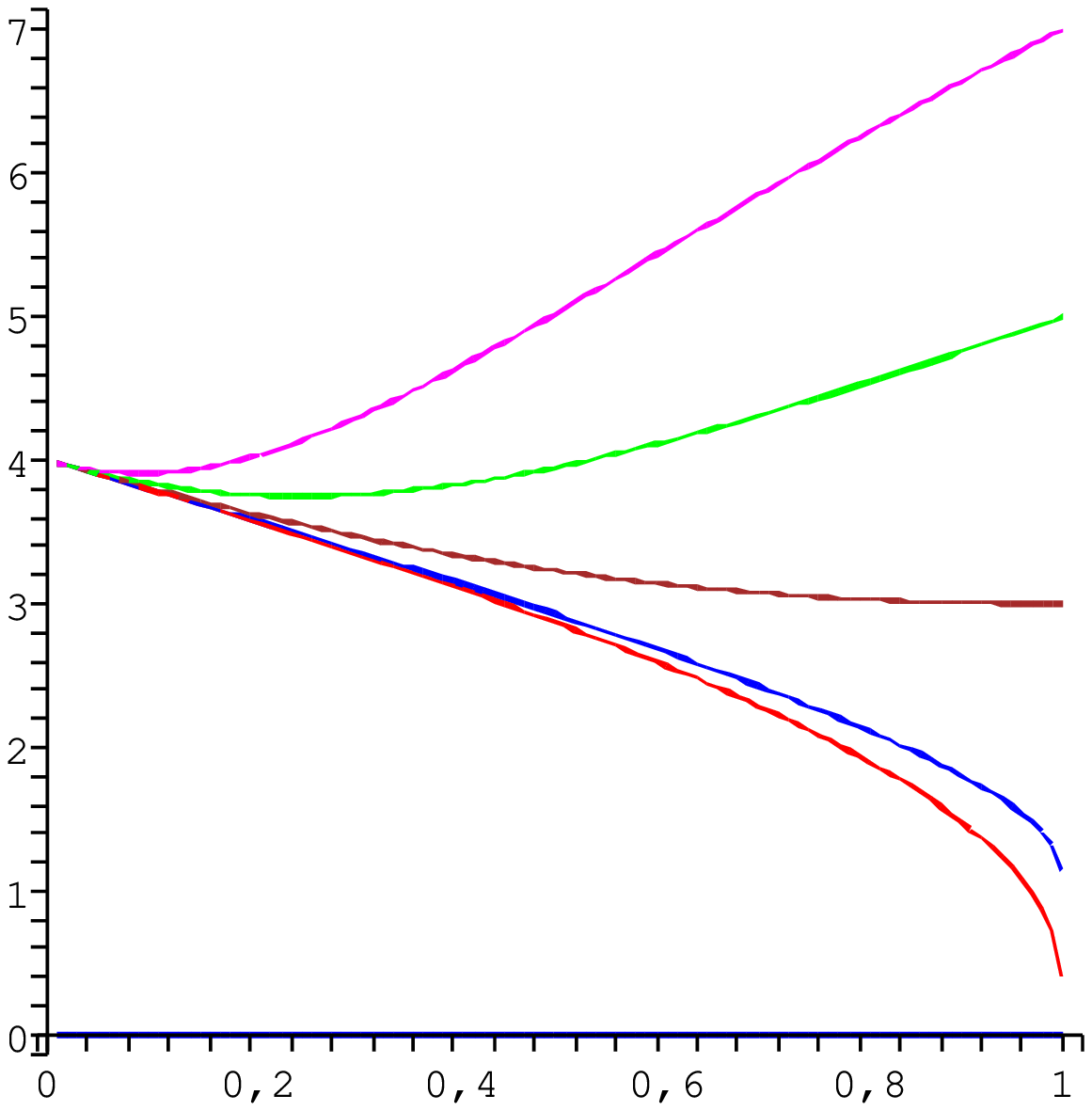}\\
      \vspace{-0.8cm} \strut
        \end{minipage}}
      \end{center}
  \vspace{-0.5cm}
  \caption{\label{fig:fig7} 
The gap in the spectrum of $\,\CH_\nu$: (a) as a function of
$\m\nu\m$ for, from top to bottom at $\nu=0$, 
$\,k^2=0,\,0.2,$\\
\hspace*{1.2cm}$0.4,\,0.6,\,0.8\,$ and $\,1$, \,(b) as a function of
$\,k^2={\alpha\over\alpha+\gamma}\,$ for, from top to bottom,
$\,\nu=4,\,3,\,2,\,2\,$ and $\,0$.}
  \vspace{0.1cm}
\end{figure}

\vskip 0.2cm  

The second approximation consisted of considering only the contributions
of the ground state of $\,\CH_\nu\,$ to the kernel of the operator
$\,\ee^{-2(\alpha+\gamma)t\m\CH_\nu}\,$ in Eq.\,(\ref{ptas1}) with
the contour integral along $\,\CC=\{Re\,\nu=\nu_{max}\}$.
\,The contributions of the excited states to that kernel decouple 
exponentially fast with a rate given by the spectral gap of 
$\,2(\alpha+\gamma)\m\CH_{\nu_{max}}\,$ for $\,\nu_{max}\,$ equal 
to the value of $\,\nu\,$ that maximizes the right hand side of 
Eq.\,(\ref{rhold}). We obtain this way a continuum of time
scales that depend on $\,\rho/t\,$ or on $\,\eta/t$. \,The plot of the 
gap $\,E_{\nu,1}-E_{\nu,0}\,$ of $\,\CH_\nu\,$ as a function of $\,\nu\,$ 
for $\,k^2=0,\,0.2,\,0.4,\,0.6,\,0.8\,$ and $\,1\,$ is given in 
Fig.\,\ref{fig:fig7}a. For $\,k^2=1\,$ the gap is expected to vanish 
for $\,-1\leq\nu\leq0\,$ and its small but positive value on the graph 
is due to the slowdown in the numerical algorithm. Indeed, as noticed 
before, the gap of $\,\CH_\nu\,$ for $\,\nu=0\,$ or $\,\nu=-1\,$ is equal 
to $\,\pi^2\bm K(k)^{-2}\,$ and tends to zero when $\,k^2\,$ tends to
$\,1\,$ (i.e. when $\,\gamma\,$ approaches $\,0$). For $\,\nu<-1\,$
or $\,\nu>0$, \,the limiting value of the gap when $\,k^2\,$
tends to zero should, however, be positive and increasing with
$\,|\nu+1/2|$. \,Fig.\,\ref{fig:fig7}b presents the gap as a function
of $\,k^2\,$ for $\,\nu=0,\,1,\,2,\,3\,$ and $\,4$. \,Since 
the interval $\,|\nu_{max}+1/2|\leq1/2\,$ corresponds to 
$\,\rho/t\leq\lambda\,$ or to $\,|\eta/t|\leq\lambda$, \,we infer that 
the decoupling of the contributions of the excited states of 
$\,\CH_\nu\,$ to the PDFs of $\,\rho\,$ or $\,\eta\,$ on those
intervals takes more and more time when $\,\gamma\to0$.
\vskip 0.2cm  

The third approximation in extracting the large deviation form of 
$\,P_t(\bm\rho)\,$ or $\,Q_t(\bm\eta)\,$ consisted in replacing the 
integral in (\ref{approxt}) along the contour 
$\,\CC=\{Re\,\nu=\nu_{max}\}$ by its saddle point value. This 
induces the correction to $\,H_\rho(\rho/t)\,$ whose leading
term comes from the one-loop contribution 
\qq
{1\over2t}\,\ln\Big({2\alpha+\beta+\gamma\,-\,(\alpha+\gamma)\,\partial_\nu^2
E_{\nu,0}|_{\nu=\nu_{max}}\over{2\alpha+\beta+\gamma\,
-\,(\alpha+\gamma)\,\partial_\nu^2E_{\nu,0}|_{\nu=0}}}\Big)\ =\ 
{1\over2t}\,\ln\left(H''(\lambda)/H''(\rho/t)\right)
\nonumber
\qqq
and similarly for $\,H_\eta(\eta/t)$. \,As we noticed at the end of the 
previous subsection, it is plausible that for $\,\wp=1\,$, 
$\,H''_\rho(\lambda)=H''_\eta(\lambda)\,$ diverges in the limit 
$\,\kappa\to1\,$ causing the divergence of the last correction.

\subsection{The case with equal Lyapunov exponents}

\noindent At the end, let us consider the special case of the potential 
flow with maximal anisotropy when $\,\wp=1\,$ and $\,\kappa=1$, i.e. when 
$\,\beta=\gamma=0$. \,In this case the strain matrix 
$\,(S^\omega_t)^i_{\,\,k}\,$ is diagonal with $\,(S^\omega_t)^1_{\,\,2}
=(S^\omega_t)^2_{\,\,1}=0\,$ and 
\qq
\Big\langle(S^\omega_t)^1_{\,\,1}\,(S^\omega_{t'})^1_{\,\,1}\Big\rangle\ =\ 
2\alpha\,\delta(t-t')\,\Big\langle(S^\omega_t)^2_{\,\,2}\,
(S^\omega_{t'})^2_{\,\,2}\Big\rangle\,,\qquad
\Big\langle(S^\omega_t)^1_{\,\,1}\,(S^\omega_{t'})^2_{\,\,2}\Big\rangle\ 
=\ 0\,.
\nonumber
\qqq
The solution of the multiplicative It\^o stochastic equation with initial
condition $\,W_0=\Id\,$ takes the form 
\qq
W_t\ =\ {\rm diag}\Big[\m\ee^{\m\int\limits_0^t 
(S^\omega_s)^1_{\,\,1}\,ds\,-\,\alpha\m t},
\,\ee^{\m\int\limits_0^t(S^\omega_s)^2_{\,\,2}\,ds\,
-\,\alpha\m t}\m\Big]\,
\nonumber
\qqq
and the stretching exponents $\,\bm\rho_t\,$ are given by the formula
\qq
{\rho_1}_{\hspace{-0.03cm}t}\,=\,
\max\Big(\ee^{\int\limits_0^t S_\omega(s)^1_{\,\,1}\,ds\,-\,\alpha\m t}\m,
\,\ee^{\int\limits_0^t S_\omega(s)^2_{\,\,2}\,ds\,
-\,\alpha\m t}\Big)\,,\quad\ 
{\rho_2}_{\hspace{-0.02cm}t}\,=\,
\min\Big(\ee^{\int\limits_0^t S_\omega(s)^1_{\,\,1}\,ds\,-\,\alpha\m t}\m,
\,\ee^{\int\limits_0^t S_\omega(s)^2_{\,\,2}\,ds\,-\,\alpha\m t}\Big)\,.\ \ 
\nonumber
\qqq
This results in the joint time $t$ PDF 
\qq
P_t(\rho_1,\rho_2)\ =\ {1\over 2\pi\alpha\m t}\m\,
\ee^{-{t\over4\alpha}\left({\rho_1\over t}+\alpha\right)^2}\,
\ee^{-{t\over4\alpha}\left({\rho_2\over t}+\alpha\right)^2}\,
\theta(\rho_1-\rho_2)
\nonumber
\qqq
or, in terms of $\,r\equiv{1\over 2}(\rho_1+\rho_2)\,$ and 
$\,\rho\equiv\rho_1-\rho_2\m$,
\qq
P_t(\rho_1,\rho_2)\ =\ {1\over 2\pi\alpha\m t}\m\,
\ee^{-{t\over2\alpha}\left({r\over t}+\alpha\right)^2}\,
\ee^{-{t\over8\alpha}\left({\rho\over t}\right)^2}\,\theta(\rho).
\nonumber
\qqq
\vskip 0.2cm

Similarly the stretching exponents $\,\bm\eta_t\,$ along the flags 
$\,F=OF^0$, are given by the Iwasawa decomposition (\ref{IW}) of the
matrices $\,\tilde W=W_tO\,$ with the flags $\,F=OF^0\,$ distributed 
with respect to the measure $\,\chi(dF)\,$ on the Grassmannian
$\,\Gr\,$ that is invariant under the process $\,W_t$, 
\,see Eq.\,(\ref{mdyc}). Any such measure has to be concentrated 
on the two flags given by the coordinate axis. On obtains then the formula
\qq
{\eta_1}_{\hspace{-0.03cm}t}\ =\ 
\ee^{\int\limits_0^t S_\omega(s)^1_{\,\,1}\,ds\,-\,\alpha\m t}\,,
\qquad{\eta_2}_{\hspace{-0.02cm}t}\ =
\ \ee^{\int\limits_0^t S_\omega(s)^2_{\,\,2}\,ds\,
-\,\alpha\m t}
\nonumber
\qqq
if $\,OF^0\,$ is given by the first coordinate axis or the one
with interchanged $\,{\eta_i}_{\hspace{-0.02cm}t}\,$
if $\,OF^0\,$ is given by the second coordinate axis. In any case,
the joint $t$ PDF of the stretching exponents $\,\bm\eta\,$ takes
the form
\qq
P_t(\eta_1,\eta_2)\ =\ {1\over 4\pi\alpha\m t}\m\,
\ee^{-{t\over4\alpha}\left({\eta_1\over t}+\alpha\right)^2}\,
\ee^{-{t\over4\alpha}\left({\eta_2\over t}+\alpha\right)^2}\,
\nonumber
\qqq
or, in terms of $\,r\equiv{1\over 2}(\eta_1+\eta_2)\,$ and 
$\,\eta\equiv\eta_1-\eta_2\m$,
\qq
P_t(\eta_1,\eta_2)\ =\ {1\over 4\pi\alpha\m t}\m\,
\ee^{-{t\over2\alpha}\left({r\over t}+\alpha\right)^2}\,
\ee^{-{t\over8\alpha}\left({\eta\over t}\right)^2}\,.
\nonumber
\qqq
\vskip 0.2cm

The PDF of $\,r_t\,$ agrees with that of Eq.\,(\ref{rld}), i.e. $r_t/t\,$
is the normal variable with mean $\,-\alpha\,$ equal to the half of the
sum of the Lyapunov exponents and with variance 
$\alpha/t$. \,As for $\,\rho_t/t\,$, \,it is distributed as an 
absolute value of the centered normal variable with variance 
$\,4\alpha/t$. \,In particular, the difference of the Lyapunov exponents 
vanishes and the large deviation rate function for $\,\rho_t\,$ 
is quadratic:
\qq
H_\rho(\rho/t)\ =\ {1\over8\alpha}\left({\rho\over t}\right)^2\,.
\label{puzz}
\qqq 
Similarly, $\,\eta/t\,$ is a normal variable with mean zero and 
variance $\,4\alpha/t\,$ and
\qq
H_\eta(\eta/t)\ =\ {1\over8\alpha}\left({\eta\over t}\right)^2\,.
\label{puze}
\qqq 
The values of the Lyapunov exponents agree with those given by the limiting 
values of Eqs.\,(\ref{slexp}) and (\ref{dlexp}). Recall however, that 
the non-Gaussianity of the large deviations, as measured by the 
difference $\,{1\over\tau}[H''_\eta(\lambda)-H''_\eta(\pm\infty)]$, 
\,was increasing with the growth of the anisotropy degree $\,\kappa$, 
\,see Fig.\,\ref{fig:fig5}. For $\,\wp=1$, \,in particular, 
$\,{1\over\tau}\m H''_\eta(\lambda)\,$ was growing with $\,\kappa\,$ whereas 
$\,{1\over\tau}\m H''_\eta(\pm\infty)={1\over2(\kappa+1)}\,$ decreased to 
the value $\,{1\over 4}\,$ for $\,\kappa=1$.  \,This seems in 
contradiction with the results (\ref{puzz}) and (\ref{puze}) with 
the quadratic large deviations rate functions for $\,\wp=1\,$ and 
$\,\kappa=1\,$ with $\,H''_\rho(\rho/t)=H''_\eta(\eta/t)={\tau\over 4}\,$ 
everywhere and not only at infinity. \,The solution of the puzzle lies 
in the non-uniformity of the large deviation regime when 
$\,\beta,\gamma\to0\,$ and the two Lyapunov exponents tend to each other. 
As we have noticed in the previous subsection, the time scales at which 
the large deviation regime sets in diverge when $\,\gamma\to0\,$ 
(and, consequently, $\,\beta\to0\,$ and $\m\lambda\to0$). 
That could explain why the limit of $\,H''_\rho(\lambda)
=H_\eta''(\lambda)\,$ when $\,\gamma\to0\,$ is not equal 
to the value of $\,H''_\rho(0)=H_\eta''(0)\,$ for $\,\gamma=0$.

\begin{figure}[ht]
\begin{center}
\vspace{0.2cm}
\mbox{\hspace{0.0cm}\psfig{file=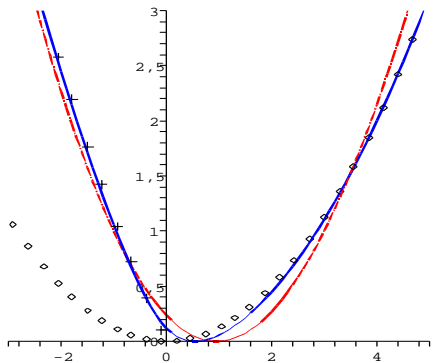,height=5.5cm,width=7cm}}
\end{center}
\vspace{-0.2cm}
\caption{Illustration of conjectured point-wise convergence of 
$\,\tau H_\eta(x/\tau)\,$ to $\,x^2/8\,$ (diamonds) for $x>0$ 
and to $\,x^2/8-x$\\
\hspace*{1.15cm}(crosses) for $x<0$. The dotted line corresponds
to $\beta=0$ and $\gamma=\alpha$, the solid one to $\beta=0$
and $\gamma=0.01\alpha$.}
\label{fig:fig8}
\vspace{-0.1cm}
\end{figure}
\vskip 0.2cm

Numerical calculations, see Fig.\,\ref{fig:fig8}, seem to indicate, 
however, that $\,H_\rho\,$ 
still converges point-wise to its form for $\,\gamma=0\,$ when $\,\gamma\to0$. 
Such point-wise convergence when $\,\gamma\to0\,$ cannot take place for 
the large deviations rate function $\,H_\eta$. \,Indeed, recall that
that for $\,\gamma>0\,$ it is $\,H_\eta(\eta/t)+\eta/(2t)\,$ that is 
an even function of $\,\eta/t\,$ whereas for $\,\gamma=0$, \,the rate function 
$\,H_\eta(\eta/t)\,$ is even itself. The point-wise convergence of 
$\,H_\eta\,$ that is a function on the whole real line cannot then hold.
Instead, for negative $\,\eta/t$, the rate function $\,H_\eta(\eta/t)\,$
should converge when $\gamma\to0\,$ to $\,H_\eta(\eta/t)-\eta/t$. 
\,Let us note that the evenness of 
$\,H_\eta(\eta/t)+\eta/(2t)\,$ for $\,\gamma>0\,$ is a consequence of 
the relations 
\qq
\CP_t(\phi,\eta)\m\equiv\m\ee^{\m t\m\CL_{\phi\eta}}
(0,0;\eta,\phi)\ =\ 
\ee^{\m\eta}\,\m\ee^{\m t\m\CL_{\phi\eta}}(\phi,\eta;0,0)\ =\ 
\ee^{\m\eta}\,\m\ee^{\m t\m\CL_{\phi\eta}}(\phi,0;0,-\eta)\,.
\label{pari}
\qqq
The first one follows from the self-adjointness of the operator 
$\,\CL_{\phi\eta}\,$ with respect to the measure 
$\,\ee^{\m\eta}\,d\phi\m d\eta\,$ 
on the product of the circle by the real line and the second one from the
commutation of $\,\CL_{\phi\eta}\,$ with the translations of $\,\eta$. 
Recall that the PDF of $\,\eta\,$ is given by the integral
\qq
\int\limits_0^\pi\ee^{\m t\m \CL_{\phi\eta}}(\phi,\eta)\,d\phi
\nonumber
\qqq
into which the initial and the final angles do not enter in a symmetric way so
that the equalities (\ref{pari}) \underline{do not} imply that
\qq
\int\limits_0^\pi\ee^{\m t\m\CL_{\phi\rho}}(0,0;\phi,\rho)\,d\phi\ =\ 
\ee^{\m\eta}\int\limits_0^\pi\ee^{\m t\m\CL_{\phi\eta}}(0,0;\phi,-\eta)
\,d\phi\,.
\nonumber
\qqq 
Nevertheless, for $\,\gamma>0\,$ the last equality holds if the full
PDF of $\,\eta\,$ are replaced by its large deviation approximation,
the angular asymmetry showing up only in the prefactors 
related to the ground state eigenfunctions of $\,\CH_\nu$. \,On the other hand,
for $\,\gamma=0\,$ the angular asymmetry does not decouple from 
the large deviation form of the PDF of $\,\eta\,$ and conspires to render 
the latter even. The lack of point-wise convergence of $\,H_\eta\,$ to its 
form for $\,\gamma=0\,$ is a reflection of the singular behavior of 
the eigenfunctions of $\,\CH_\nu\,$ when $\,\gamma\to0$.  
\vskip 0.3cm

To summarize, although the distribution of the stretching exponents 
$\,\bm\eta\,$ still exhibits large deviation regime when $\,\gamma=0$, 
\,the corresponding rate function is not equal to the limit of the rate 
functions for $\,\gamma>0\,$ signaling that when two Lyapunov exponents 
coincide, the occurrence of the multiplicative large deviation regime 
becomes problematic.

\nsection{Conclusions}

We have examined in detail the tangent process $\,W^\omega_t\,$ describing 
the evolution of infinitesimal separations between Lagrangian trajectories 
in the two-dimensional Kraichnan flow in a periodic square. The process 
$\,W^\omega_t\,$ is driven by the time decorrelated strain whose distribution 
is, in general, anisotropic, possessing only the symmetry with respect of the 
$\,90^o\,$ rotations and axes reflections. Our interest was concentrated
on the large deviation regime of the stretching exponents $\,\rho_i\,$
or $\,\eta_i\,$ that appear in the matrix decompositions 
$\,W=O'\,{\rm diag}[\ee^{\m\rho_1},\ee^{\m\rho_2}]\,O\,$ or 
$\,W=O'\,{\rm diag}[\ee^{\m\eta_1},\ee^{\m\eta_2}]\,N\,$ with orthogonal 
matrices $\,O,O'\,$ and upper-triangular $\,N$. 
\,The anisotropy couples the dynamics of the  stretching exponents to 
the evolution of the matrices $\,O'$, \,in contrast to the situation in 
the isotropic case where the stochastic dynamics of the stretching 
exponents is decoupled from that of the matrices $\,O'\m$ and $\,O\,$ 
or $\,O'\m$ and $\,N$. \,The stochastic evolution of the matrices $\,O'\,$ 
becomes, however, independent, at least at long times, from that of the 
stretching exponents, attaining exponentially fast a stationary state. The 
latter feeds to the evolution of the stretching exponents in a steady 
fashion permitting them still to attain the large deviation regime. The 
large deviation rate function for the stretching exponents may be expressed 
in terms of the ground state energy of an operator on the group of 
orthogonal matrices parametrized by the variables conjugate to the 
stretching exponents. The contribution of the excited states to the
PDF of the stretching exponents decouples exponentially fast with the rate
equal to the gap of the angular operator. This scenario for the 
multiplicative large deviations seems quite general whenever the Lyapunov 
exponents are all different, at least in the homogeneous Kraichnan model. 
When some of the Lyapunov exponents become close, some of the time scales 
for the appearance of the large deviation regime as well as the prefactors 
multiplying the exponential large deviation PDF may diverge. 
\vskip 0.2cm

What is special about the flow on the periodic square is that 
the operator on the orthogonal group in question takes form of 
the integrable periodic Schr\"odinger operator of the Lam\'e type 
facilitating the calculations. This simplification due to the 
hidden integrable structure allowed to obtain closed formulae 
for the Lyapunov exponents in terms of elliptic integrals and to analyze 
the large deviation rate function with precision. The results of 
the analysis show that the anisotropy effects lower the top Lyapunov 
exponent and increase the lower one (relative to an overall inverse 
time scale), with the sum of the two fixed by the compressibility
degree of the flow. The sum of the two stretching exponents 
is normally distributed with the covariance again fixed by the 
compressibility degree. The difference of the stretching exponents, 
however, exhibits in the presence of anisotropy non-Gaussian large 
deviations. Its central limit covariance grows with anisotropy but 
the quadratic large-value asymptotes of the rate function have the top 
coefficients that decrease with increasing anisotropy. This non-Gaussian 
scenario for the multiplicative large deviations applies, however, only 
when the Lyapunov exponents are different. At the extreme anisotropy, 
when the strain matrix is diagonal with independent equally distributed 
entries, the two Lyapunov exponents coincide and the large deviations 
for the stretching exponents are Gaussian. We have analyzed in more 
detail this discontinuous restoration of the Gaussianity of large 
deviations. 
\vskip 0.2cm

For the Kraichnan flow in a periodic rectangle, the multiplicative 
large deviations may be analyzed similarly. The results will be published  
elsewhere. It remains an open question whether the Kraichnan flow   
in a three-dimensional periodic box possesses a hidden integrable 
structure that would permit to extend the analysis of the present paper
to that case.

\end{document}